\documentclass[preprint,aps,tightenlines,byrevtex,showpacs,nofootinbib]{revtex4}
\usepackage{graphicx}
\begin{document}

\title{On nonadiabatic contributions
to the neutrino oscillation probability
and the formalism by Kimura, Takamura and Yokomakura}

\author{Osamu Yasuda}
\affiliation{Department of Physics, Tokyo Metropolitan University,
Minami-Osawa, Hachioji, Tokyo 192-0397, Japan}

\begin{abstract}

It is shown that it is possible to obtain
the analytical expression for the effective mixing angle in matter
using the formalism which was developed by Kimura, Takamura and
Yokomakura for the neutrino oscillation probability in matter
with constant density.
If we assume that the imaginary part of
the integral of the difference of the energy eigenvalues of the
two levels at each level-crossing is given by the ratio $\gamma$ of
the difference of the energy eigenvalues of the two levels to
the derivative of the effective mixing angle
at the level-crossing, then the nonadiabatic contribution
to the oscillation probability can be expressed analytically
by this formalism.
We give one example in which the energy eigenvalues
cannot be expressed as roots of a quadratic equation and
we show that our assumption is correct in the 
approximation of the small mixing angle.

\end{abstract}
\vskip 0.1cm
\maketitle

\section{introduction}

Since the discovery of oscillation of atmospheric neutrinos
\cite{Fukuda:1998mi}, neutrino oscillation has attracted a lot of
attention.  To discuss the behaviors of neutrino oscillation
intuitively, it is important to have analytical formulae for the
oscillation probability.  However, it is difficult to obtain an
analytical formula in the three flavor mixing scheme in matter.  In
2002 Kimura, Takamura and Yokomakura (KTY) discovered a compact
formula~\cite{Kimura:2002hb,Kimura:2002wd} for the neutrino
oscillation probability in matter with constant density.  Subsequently
the KTY framework was
generalized to more general cases.  Ref.\,\cite{Zhang:2006yq} discussed
the four neutrino mixing scheme in matter with constant density.
Ref.\,\cite{Yasuda:2007jp} discussed two cases of neutrino oscillation
in the adiabatic approximation,
the one with non-standard interactions where
the matter potential has non-diagonal elements in the flavor basis, or
the other with large neutrino magnetic moments in a magnetic field.

In general, however, adiabatic approximation may not be good, and in
the present paper we discuss nonadiabatic contributions to the
oscillation probability.
It is believed\,\footnote{See, e.g., Ref.\,\cite{Kuo:1989qe}
and references therein. See also Ref.\,\cite{Yamamoto:2010jc}
for a discussion on the condition to justify such a
treatment.}
that the nonadiabatic
contributions to the transition phenomena in a problem with three or
more eigenstates can be treated approximately well by applying the
method for two state problems
\cite{landau,Zener:1932ws} at each level-crossing, if the
the two resonances are sufficiently far apart.

In the two flavor case, nonadiabatic contributions to the oscillation
probability is
approximately obtained by the WKB method~\cite{landau} in which
the imaginary part of
the integral of the difference of the energy eigenvalues of the
two levels is evaluated.
The imaginary part of this integral
is proportional to the ratio $\gamma$ of the difference
of the energy eigenvalues of the two levels to
the derivative of the effective mixing angle
at the level-crossing.
In the three flavor case with standard matter effect,
the situation at each level-crossing is essentially
the same as that in the two flavor case, because
the energy eigenvalues are the roots of quadratic
equations and the procedure for diagonalization
is the same as that for the two flavor case.
So one can discuss analytically the oscillation probability of supernova
neutrinos\,\footnote{
It was pointed out that
so-called collective
oscillations\,\cite{Pantaleone:1992eq,Samuel:1993uw}
could be important for phenomenology of supernova neutrinos.
However, this topics is beyond the scope of paper,
and we do not discuss the effect of collective
oscillations here.}, which are supposed to
go through the two level-crossing points.

In general case of neutrino oscillation with nonstandard matter
effects or with more than three flavors, however, it is nontrivial
to obtain the analytical expression for the energy eigenvalues
and the effective mixing angles.
In this paper we show that, if the energy eigenvalues
are obtained analytically, then it is possible to obtain
the analytical expression for the effective mixing angle
using the KTY formalism.  Furthermore, assuming that the imaginary part of
the integral of the difference of the energy eigenvalues of the
two levels at each level-crossing is given by the ratio $\gamma$ of
the difference of the energy eigenvalues of the two levels to
the derivative of the effective mixing angle
at the level-crossing, we argue that nonadiabatic contributions
to the oscillation probability can be expressed analytically
by the KTY formalism.
We give one example in which the energy eigenvalues
cannot be expressed as roots of a quadratic equation and
we show that our assumption is correct in the small
mixing angle limit.

Throughout this paper we discuss the case in which
the baseline of the neutrino path is long enough
so that averaging over rapid oscillations
is a good approximation,
as in the case of the solar neutrino deficit phenomena.
We also assume normal hierarchy
for simplicity.  The case with inverted hierarchy can
be treated by the same manner.

The paper is organized as follows. In Sec. II, we review
basic results of neutrino oscillation
in the adiabatic approximation.
In Sec. III, we discuss the nonadiabatic contributions
to the flavor transition, and show how to express
the effective mixing angle using the KTY formalism.
In Sec. IV, we apply our general idea of Sec. III to
two examples.  In Sec. V, we draw our conclusions.
In the appendices A, B, C, D, E, F, G , and H we
provide details of the derivation of
the analytic formulae for the energy eigenvalues,
the effective mixing angles and the nonadiabatic
contributions to oscillation probabilities.

\section{The oscillation probability in the adiabatic
approximation and the effective mixing angles in matter}

\subsection{The oscillation probability in matter}
The equation of motion for neutrinos propagating in matter with
general potential is given by
\begin{eqnarray}
i{d\Psi \over dt}=
\left[U{\cal E}_0U^{-1}
+{\cal A}(t)
\right]\Psi,
\label{sch1}
\end{eqnarray}
where
\begin{eqnarray}
{\cal E}_0&\equiv&{\mbox{\rm diag}}\left(E_1,E_2,E_3\right),
\nonumber\\
{\cal A}(t)&\equiv&
\left(
\begin{array}{ccc}
 A_{ee}(t) & A_{e\mu}(t) & A_{e\tau}(t)\\
 A_{\mu e}(t) & A_{\mu\mu}(t) & A_{\mu\tau}(t)\\
 A_{\tau e}(t) & A_{^\ast\mu}(t) & A_{\tau\tau}(t)
\end{array}
\right).
\nonumber
\end{eqnarray}
$\Psi^T\equiv(\nu_e,\nu_\mu,\nu_\tau)$ is the flavor eigenstate, $U$
is the leptonic mixing matrix, $E_j\equiv\sqrt{m_j^2 +{\vec
{\relax{\kern .1em p}}}^2}~(j=1,2,3)$ is the energy eigenvalue of each
mass eigenstate, and the matter effect ${\cal A}(t)$ at time (or
position ) $t$ is characterized by the matter potential
$A_{\alpha\beta}(t)~(\alpha,\beta=e,\mu,\tau)$.
Since the matrix which is proportional to identity gives
contribution only to the phase of the probability amplitude,
instead of ${\cal E}_0$ itself, we use the following quantity:
\begin{eqnarray}
{\cal E}\equiv{\cal E}_0-E_1\mbox{\bf 1}=\mbox{\rm diag}(0,\Delta E_{21},\Delta
E_{31}),
\nonumber
\end{eqnarray}
where 
\begin{eqnarray}
\Delta E_{jk}\equiv E_j - E_k \simeq (m_j^2-m_k^2)/2
|{\vec p}|.
\nonumber
\end{eqnarray}
The $3\times3$ matrix on the
right hand side of Eq.\,(\ref{sch1}) can be formally
diagonalized as:
\begin{eqnarray}
U{\cal E}U^{-1}+{\cal A}(t)
=\tilde{U}(t)\tilde{{\cal E}}(t)\tilde{U}^{-1}(t),
\label{sch3}
\end{eqnarray}
where
\begin{eqnarray}
\tilde{{\cal E}}(t)&\equiv&{\mbox{\rm diag}}\left(
\tilde{E}_1(t),\tilde{E}_2(t),\tilde{E}_3(t)\right)
\nonumber
\end{eqnarray}
is a diagonal matrix with the energy eigenvalues
$\tilde{E}_j(t)$ in the presence of the matter effect.

In this section we consider the case where the density of
the matter varies adiabatically as in the case of
the solar neutrino deficit phenomena.
In this case, we get
\begin{eqnarray}
\Psi(L)=\tilde{U}(L)\exp\left[-i\int_0^L\tilde{{\cal E}}(t)\,dt\right]
\tilde{U}(0)^{-1}\Psi(0),
\label{psi1}
\end{eqnarray}
where $\tilde{U}(0)$ and $\tilde{U}(L)$ stand for
the effective mixing matrices at the origin $t=0$ and
at the endpoint $t=L$.
The oscillation probability is given by
\begin{eqnarray}
P(\nu_\alpha\rightarrow\nu_\beta)&=&
\sum_{j,k}\,\tilde{U}_{\beta j}(L)\tilde{U}^\ast_{\beta k}(L)
\tilde{U}^\ast_{\alpha j}(0)\tilde{U}_{\alpha k}(0)
\exp\left[-i\int_0^L\Delta\tilde{E}_{jk}(t)\,dt\right],
\label{proba1}
\end{eqnarray}
where we have defined
\begin{eqnarray}
\Delta \tilde{E}_{jk}(t)&\equiv&\tilde{E}_j(t)-\tilde{E}_k(t).
\nonumber
\end{eqnarray}
Eq.\,(\ref{proba1}) requires
the quantity $\tilde{U}_{\beta j}(t)\tilde{U}^\ast(t)_{\beta k}$
which has the same flavor index $\beta$
but different mass eigenstate indices $j, k$,
and it turns out that the analytical expression for
$\tilde{U}_{\beta j}(t)\tilde{U}^\ast(t)_{\beta k}$
is hard to obtain.  However,
if the length $L$ of the neutrino path is very large
and if $|\int_0^L\Delta\tilde{E}(t)_{jk}\,dt|\gg 1$ is satisfied
for $j\ne k$, then, after averaging over rapid oscillations
as in the case of the solar neutrino deficit phenomena,
Eq.\,(\ref{proba1}) is reduced to
\begin{eqnarray}
P(\nu_\alpha\rightarrow\nu_\beta)
=\sum_j\,\tilde{X}^{\beta\beta}_j(L)\tilde{X}^{\alpha\alpha}_j(0)
=\sum_j\,X^{\beta\beta}_j\tilde{X}^{\alpha\alpha}_j(0)
\label{adiabatic}
\end{eqnarray}
where we have defined
\begin{eqnarray}
\tilde{X}^{\alpha\beta}_j(t)&\equiv&
\tilde{U}_{\alpha j}(t)\tilde{U}_{\alpha j}(t)^\ast,
\label{tildex}\\
\tilde{X}^{\alpha\alpha}_j(t)&\equiv&
\left|\tilde{U}_{\alpha j}(t)\right|^2.
\nonumber
\end{eqnarray}
and here and in the following we assume that
there is no matter at the end of the baseline $t=L$.

It is known\,\cite{Kimura:2002hb,Kimura:2002wd,Xing:2005gk,Yasuda:2007jp}
that the quantity
$\tilde{X}^{\alpha\beta}_j(t)\equiv
\tilde{U}_{\alpha j}(t)\tilde{U}_{\alpha j}(t)^\ast$
can be expressed as

\begin{eqnarray}
\left(\begin{array}{c}
\tilde{X}^{\alpha\beta}_1\cr
\tilde{X}^{\alpha\beta}_2\cr
\tilde{X}^{\alpha\beta}_3
\end{array}\right)
=\left(\begin{array}{ccc}
\displaystyle
\frac{{\ }1}{\Delta \tilde{E}_{21} \Delta \tilde{E}_{31}}
(\tilde{E}_2\tilde{E}_3 & -(\tilde{E}_2+\tilde{E}_3)&
1)\cr
\displaystyle
\frac{-1}{\Delta \tilde{E}_{21} \Delta \tilde{E}_{32}}
(\tilde{E}_3\tilde{E}_1 & -(\tilde{E}_3+\tilde{E}_1)&
1)\cr
\displaystyle
\frac{{\ }1}{\Delta \tilde{E}_{31} \Delta \tilde{E}_{32}}
(\tilde{E}_1\tilde{E}_2 & -(\tilde{E}_1+\tilde{E}_2)&
1)\cr
\end{array}\right)
\left(\begin{array}{r}
\delta_{\alpha\beta}\cr
\left[U{\cal E}U^{-1}+{\cal A}\right]_{\alpha\beta}\cr
\left[\left(U{\cal E}U^{-1}+{\cal A}\right)^2\right]_{\alpha\beta}
\end{array}\right),
\label{solx}
\end{eqnarray}
where the $t-$dependence of
the quantities $\tilde{X}^{\alpha\beta}_j$,
$\tilde{E}_j$, $\Delta \tilde{E}_{jk}$
is suppressed for simplicity in Eq.\,(\ref{solx}).
$[(U{\cal E}U^{-1}+{\cal A})^j]_{\alpha\beta}$ $(j=1,2)$ on
the right hand side are given by the known quantities
although the computations are tedious for general potential
${\cal A}(t)$.

\section{the nonadiabatic corrections to the
oscillation probability\label{nonadiabatic}}

\subsection{The  the standard case}

When adiabatic approximation is not good,
Eq.\,(\ref{adiabatic}) should be modified by taking
nonadiabatic contributions into account.
Substituting the diagonalized form (\ref{sch3})
of the Hamiltonian into the Dirac equation (\ref{sch1}),
we have
\begin{eqnarray}
i\frac{d{\ }}{dt}\left\{
\tilde{U}^{-1}(t)\Psi(t)\right\}
=
\left[\tilde{\cal E}-i\tilde{U}^{-1}
\left(\frac{d{\ }}{dt}\tilde{U}\right)
\right]\left\{\tilde{U}^{-1}(t)\Psi(t)\right\},
\label{sch4}
\end{eqnarray}
where $\tilde{U}^{-1}(t)\Psi(t)$ is the effective energy
eigenstate.

In the two flavor case, there is only
one level crossing, and the probability
$P(\nu_\alpha\to\nu_\beta)$ is given by
\begin{eqnarray}
P(\nu_\alpha\to\nu_\beta)
&=&\left(\begin{array}{cc}
\left|U_{\beta 1}\right|^2&
\left|U_{\beta 2}\right|^2
\end{array}\right)
\left(\begin{array}{cc}
1-P_C & P_C\\
P_C & 1-P_C
\end{array}\right)
\left(\begin{array}{c}
\left|\tilde{U}_{\alpha 1}(0)\right|^2\\
\left|\tilde{U}_{\alpha 2}(0)\right|^2
\end{array}\right),
\nonumber
\end{eqnarray}
where $P_C$ stands for the jumping
probability from the energy eigenstate
$\tilde{\nu}_1$ to $\tilde{\nu}_2$, and is
approximately
computed by the WKB method\,\cite{landau}:
\begin{eqnarray}
P_C= \exp\left[ -\,\mbox{\rm Im}\int_C\,
\Delta \tilde{E}(t)\,dt
\right].
\label{pc}
\end{eqnarray}
Here $\Delta \tilde{E}(t)$ is the difference of the two energy
eigenvalues at the level crossing, and the contour $C$
is defined as a path from $t=t_0$ to $t=t_1$ where
$t_0$ is a point which gives the minimum value of $\Delta
\tilde{E}(t)$,
and $t_1$ is a point in the complex $x$-plane
such that $\Delta\tilde{E}(t_1)=0$.
Throughout this paper we assume
that the WKB approximation (\ref{pc}) is good\,\footnote{
In the extreme nonadiabatic limit, Eq.\,(\ref{pc})
need a certain modification\,\cite{Kuo:1988pn}.
However, the major purpose of this paper is to show
how to handle nonadiabatic corrections in the case
with more than two flavors, rather than computing the
deviation from the linear potential or the
corrections due to the extreme nonadiabatic condition.
These corrections can be treated in the same manner
as in the two flavor case, so we do not include such a modification
for simplicity in this paper.}.
It is known that the exponent in (\ref{pc})
is related by the ratio $\gamma$ of
the difference of the energy eigenvalues of the two levels to
the derivative of the effective mixing angle $\tilde{\theta}$
at the level-crossing:
\begin{eqnarray}
-\log P_C = \frac{\pi}{2}\,F\,\gamma,
\label{logpc}
\end{eqnarray}
where
\begin{eqnarray}
\gamma = \left.
\displaystyle\frac{\Delta \tilde{E}}
{2
|
d\tilde{\theta}/
dt
|}
\right|_{\mbox{\rm\scriptsize resonance}}.
\nonumber
\end{eqnarray}
The subscript resonance stands for the quantity
evaluated at the point where the difference
$|\Delta \tilde{E}|$
of the two energy eigenvalues becomes minimum.
$F$ is the factor which depends on the
form of the potential $A$, and in the case of
linear potential, i.e., in the case $A\propto t$
we have $F=1$\,\footnote{
$F$ can be evaluated analytically
for a few cases of the density profile such as
$t$, $1/t$ and $e^{-t}$.
See Ref.\,\cite{Kuo:1989qe} and references therein.}.
In the standard two flavor case, we have
\begin{eqnarray}
\gamma = 
\displaystyle\frac{\Delta E_{21}\sin^22\theta}
{\cos2\theta
\left
|
d\log A/
dt
\right|_{\mbox{\rm\scriptsize resonance}}}.
\nonumber
\end{eqnarray}

In the three flavor case, there are
at most two level crossings for neutrinos as in the case
of a supernova\,\cite{Dighe:1999bi}.
Assuming that the nonadiabatic transition
at each level crossing can be described
by the same manner as in the case of
the two level-crossing, and
assuming that the level crossing occurs at
$\tilde{E}_3\simeq\tilde{E}_1$ and
$\tilde{E}_2\simeq\tilde{E}_1$,
the transition probability is given by
\begin{eqnarray}
P(\nu_\alpha\to\nu_\beta)
&=&\left(\begin{array}{ccc}
\left|U_{\beta 1}\right|^2&
\left|U_{\beta 2}\right|^2&
\left|U_{\beta 3}\right|^2
\end{array}\right)
\left(\begin{array}{ccc}
1-P_L & P_L& 0 \\
 P_L  & 1-P_L & 0\\
0 &0 & 1
\end{array}\right)\nonumber\\
&{\ }&\times
\left(\begin{array}{ccc}
1-P_H & 0 & P_H\\
0 & 1 & 0\\
P_H & 0 & 1-P_H
\end{array}\right)
\left(\begin{array}{c}
\left|\tilde{U}_{\alpha 1}(0)\right|^2\\
\left|\tilde{U}_{\alpha 2}(0)\right|^2\\
\left|\tilde{U}_{\alpha 3}(0)\right|^2
\end{array}\right).
\label{p3}
\end{eqnarray}
Using the WKB approximation\,\cite{landau},
the jumping factors in (\ref{p3}) are given by
\begin{eqnarray}
P_H&=& \exp\left[ -\,\mbox{\rm Im}\int_C\,
\Delta \tilde{E}_{31}(t)\,dt
\right],
\nonumber\\
P_L&=& \exp\left[ -\,\mbox{\rm Im}\int_C\,
\Delta \tilde{E}_{21}(t)\,dt
\right].
\nonumber
\end{eqnarray}

As is shown in Appendix \ref{appendixa},
\begin{eqnarray}
\tilde{\theta}_{13}&\equiv&\frac{1}{2}
\tan^{-1}
\frac{\Delta E_{31}\sin2\theta_{13}}
{\Delta E_{31}\cos2\theta_{13}-A}
\label{th13t}
\end{eqnarray}
is the effective mixing angle
near the level-crossing
$\Delta E_{31}\simeq A$,
while
\begin{eqnarray}
\tilde{\theta}_{12}&\equiv&\frac{1}{2}
\tan^{-1}
\frac{\Delta E_{21}\sin2\theta_{12}}
{\Delta E_{21}\cos2\theta_{12}-Ac^2_{13}}
\nonumber
\end{eqnarray}
is the effective mixing angle
near the level-crossing
$\Delta E_{21}\simeq A$.
As in the case of a two level
problem, we have the following
jumping probabilities:
\begin{eqnarray}
P_H&=& 
\displaystyle\exp\left(-\frac{\pi}{2}\,F\,
\frac{\Delta E_{31}\sin^22\theta_{13}}
{\cos2\theta_{13}
\left
|
d\log A/
dt
\right|_{\mbox{\rm\scriptsize resonance}}}\right),
\label{ph}\\
P_L&=& 
\displaystyle\exp\left(-\frac{\pi}{2}\,F\,
\frac{\Delta E_{21}\sin^22\theta_{12}}
{\cos2\theta_{12}
\left
|
d\log A/
dt
\right|_{\mbox{\rm\scriptsize resonance}}}\right).
\label{pl}
\end{eqnarray}
$F$ is the factor which depends on the
form of the potential $A$, and $F=1$ in the case of
a linear potential.

The effective mixing matrix elements
$|\tilde{U}_{\alpha j}(0)|^2$ at the origin $L=0$
can be approximately obtained by putting
$\Delta m^2_{21}\to 0$ and by
substituting $\theta_{12}\to 0$,
$\theta_{13}\to \tilde{\theta}_{13}$,
$\delta\to 0$. They are thus given by
\begin{eqnarray}
|\tilde{U}_{\alpha j}(0)|^2&=&\left(
\begin{array}{ccc}
\tilde{c}_{13}^2&0&\tilde{s}_{13}^2\cr
\tilde{s}_{13}^2s_{23}^2&c_{23}^2&\tilde{c}_{13}^2s_{23}^2\cr
\tilde{s}_{13}^2c_{23}^2&s_{23}^2&\tilde{c}_{13}^2c_{23}^2
\end{array}\right),
\label{u02}
\end{eqnarray}
where $\tilde{c}_{13}\equiv\cos\tilde{\theta}_{13}$ and
$\tilde{s}_{13}\equiv\sin\tilde{\theta}_{13}$ are
the quantities which are evaluated from Eq.\,(\ref{th13t})
at the origin $L=0$.
From Eqs.\,(\ref{p3}), (\ref{ph}), (\ref{pl}) and (\ref{u02}), 
we have the analytical expression for the
transition probability including the
nonadiabatic contributions in the three flavor standard case.

\subsection{The effective mixing angles in matter}
The energy eigenvalues in the previous discussions
are easily obtained because the
eigenvalues near each level-crossing in the standard case
are the roots of a quadratic equation
in the leading order in $\Delta E_{21}/\Delta E_{31}$.
In some cases with more than two states, however,
the energy eigenvalues cannot be expressed as
the roots of a quadratic equation, and in that case
it is difficult to compute the jumping factor (\ref{pc}).
Assuming that the equality (\ref{logpc})
holds at each level-crossing,
it is useful
to define the effective mixing angle in the presence
of the matter potential ${\cal A}(t)$\,\footnote{
The effective mixing angles were given in the standard three flavor
case in Ref.\,\cite{Zaglauer:1988gz}}.

In this subsection we will show how to derive
the expression for the effective mixing angle in the presence
of the matter using the KTY formalism.
The KTY formalism
enables one to obtain the bilinear quantity
$\tilde{X}^{\alpha\beta}_j$ defined in Eq.\,(\ref{tildex}).
Our strategy here is to start with effective matrix elements
$\tilde{X}^{\alpha\beta}_j$ which are obtained by the KTY formalism
and to determine the phase of each element by demanding
that it be consistent with the standard form (\ref{param1})
of the mixing matrix element in vacuum.
We will discuss only the three flavor case, but we can
generalize this method to the case with more than three flavors.

From the identities
\begin{eqnarray}
\tilde{U}_{ej}&=&\sqrt{\tilde{X}^{ee}_j}\,e^{i\arg{\tilde{U}_{ej}}},
\nonumber\\
\tilde{U}_{\mu j}&=&\frac{\tilde{X}^{\mu e}_j}
{\sqrt{\tilde{X}^{ee}_j}}\,e^{i\arg{\tilde{U}_{ej}}},
\nonumber\\
\tilde{U}_{\tau j}&=&\frac{\tilde{X}^{\tau e}_j}
{\sqrt{\tilde{X}^{ee}_j}}\,e^{i\arg{\tilde{U}_{ej}}},
\nonumber
\end{eqnarray}
the first guess for the effective mixing matrix $\tilde{U}$
can be written as
\begin{eqnarray}
\tilde{U}_1=\tilde{U}_0\times\mbox{\rm diag}\left(
e^{i\arg{\tilde{U}_{e1}}},
e^{i\arg{\tilde{U}_{e2}}},
e^{i\arg{\tilde{U}_{e3}}}
\right),
\label{u1}
\end{eqnarray}
where
\begin{eqnarray}
\tilde{U}_0\equiv
\left(\begin{array}{ccc}
\left(\begin{array}{c}
\sqrt{\tilde{X}^{ee}_1}\cr
\tilde{X}^{\mu e}_1
/\sqrt{\tilde{X}^{ee}_1}
\cr
\tilde{X}^{\tau e}_1
/\sqrt{\tilde{X}^{ee}_1}
\end{array}\right),&
\left(\begin{array}{c}
\sqrt{\tilde{X}^{ee}_2}\cr
\tilde{X}^{\mu e}_2
/\sqrt{\tilde{X}^{ee}_2}\cr
\tilde{X}^{\tau e}_2
/\sqrt{\tilde{X}^{ee}_2}
\end{array}\right),&
\left(\begin{array}{c}
\sqrt{\tilde{X}^{ee}_3}\cr
\tilde{X}^{\mu e}_3
/\sqrt{\tilde{X}^{ee}_3}\cr
\tilde{X}^{\tau e}_3
/\sqrt{\tilde{X}^{ee}_3}
\end{array}\right)
\end{array}\right).
\label{u0}
\end{eqnarray}
However, this naive choice is not
exactly the same as the standard parametrization (\ref{param1})
for the vacuum mixing matrix,
because Eq.\,(\ref{param1}) implies
that Im($\tilde{U}_{e1}$)=Im($\tilde{U}_{e2}$)=
Im($\tilde{U}_{\mu3}$)=Im($\tilde{U}_{\tau3}$)=0
and det$\tilde{U}$=1.
In order for $\tilde{U}_{\alpha j}$ to have the expression
consistent with the form (\ref{param1}), therefore,
we postulate the following conditions:
\begin{eqnarray}
\arg{\tilde{U}_{e1}}&=&0,
\label{phasee1}\\
\arg{\tilde{U}_{e2}}&=&0,
\label{phasee2}\\
\arg{\tilde{U}_{\mu3}}&=&0,
\label{phasem3}\\
\arg{\tilde{U}_{\tau3}}&=&0,
\label{phaset3}\\
\arg\det{\tilde{U}_{\alpha j}}&=&0.
\label{phasedet}
\end{eqnarray}
To satisfy Eqs.\,(\ref{phasee1})--(\ref{phasedet}),
we multiply diagonal matrices with elements with
complex phases both from the left and right
hand sides of $\tilde{U}_0$\,\footnote{
Since we are using all the available degrees of freedom
of the phases in the 3$\times$3 matrix,
it does not matter whether we start with $\tilde{U}_0$
or $\tilde{U}_1$.  For simplicity we start
with $\tilde{U}_0$ here.}:
\begin{eqnarray}
\tilde{U}\equiv e^{i\varphi_0}\,e^{i\varphi_3\lambda_3}\,
e^{i\varphi_9\lambda_9}\,\tilde{U}_0\,
e^{i\varphi'_9\lambda_9}\,e^{i\varphi'_3\lambda_3},
\label{utilde1}
\end{eqnarray}
where
\begin{eqnarray}
\lambda_3&\equiv&\mbox{\rm diag}(1,-1,0),
\label{lambda3}\\
\lambda_9&\equiv&\mbox{\rm diag}(1,0,-1).
\label{lambda9}
\end{eqnarray}
It is straightforward to obtain
$\varphi_0$, $\varphi_3$, $\varphi_9$, $\varphi'_3$, $\varphi'_9$
from Eqs.\,(\ref{phasee1})--(\ref{phasedet}),
and we get
\begin{eqnarray}
\varphi_0&=&-\frac{1}{3}\arg\det\tilde{U}_0,
\label{alpha}\\
\varphi_3&=&~~\,\frac{1}{3}\arg\det\tilde{U}_0
-\frac{1}{3}\arg\tilde{X}^{\mu e}_1
+\frac{1}{3}\arg\tilde{X}^{\mu e}_3
-\frac{2}{3}\arg\tilde{X}^{\tau e}_3,
\label{beta}\\
\varphi_9&=&~~\,\frac{1}{3}\arg\det\tilde{U}_0
-\frac{1}{3}\arg\tilde{X}^{\mu e}_1
-\frac{2}{3}\arg\tilde{X}^{\mu e}_3
+\frac{1}{3}\arg\tilde{X}^{\tau e}_3,
\label{betap}\\
\varphi'_3&=&~~\,\frac{1}{3}\arg\det\tilde{U}_0
+\frac{1}{3}\arg\tilde{X}^{\mu e}_1
-\frac{1}{3}\arg\tilde{X}^{\mu e}_3
-\frac{1}{3}\arg\tilde{X}^{\tau e}_3,
\label{gamma}\\
\varphi'_9&=&-\frac{2}{3}\arg\det\tilde{U}_0
+\frac{1}{3}\arg\tilde{X}^{\mu e}_1
+\frac{2}{3}\arg\tilde{X}^{\mu e}_3
+\frac{2}{3}\arg\tilde{X}^{\tau e}_3.
\label{gammap}
\end{eqnarray}
With the values in Eqs.\,(\ref{alpha})--(\ref{gammap}),
$\tilde{U}$ in Eq.\,(\ref{utilde1}) has the same
parametrization as that for the standard one (\ref{param1}):
\begin{eqnarray}
\tilde{U}=e^{i\tilde{\theta}_{23}\lambda_7}\,
\Gamma_{\tilde{\delta}}^{(13)}\,
e^{i\tilde{\theta}_{13}\lambda_5}\,
(\Gamma_{\tilde{\delta}}^{(13)})^{-1}\,
e^{i\tilde{\theta}_{12}\lambda_2},
\label{utilde2}
\end{eqnarray}
where $\lambda_j~(j=2,5,7)$ are
the Gell-Mann matrices and defined by
\begin{eqnarray}
\lambda_2&\equiv&
\left(\begin{array}{ccc}
0&-i&0\cr
i&0&0\cr
0&0&0
\end{array}\right),
\label{lambda2}\\
\lambda_5&\equiv&
\left(\begin{array}{ccc}
0&0&-i\cr
0&0&0\cr
i&0&0
\end{array}\right),
\label{lambda5}\\
\lambda_7&\equiv&
\left(\begin{array}{ccc}
0&0&0\cr
0&0&-i\cr
0&i&0
\end{array}\right).
\label{lambda7}
\end{eqnarray}
Comparing Eqs.\,(\ref{u0}), (\ref{utilde1}) and (\ref{utilde2}),
we find
\begin{eqnarray}
\cos2\tilde{\theta}_{12}&=&\frac
{\tilde{X}^{ee}_1-\tilde{X}^{ee}_2}
{\tilde{X}^{ee}_1+\tilde{X}^{ee}_2},
\label{theta12tilde}\\
\cos2\tilde{\theta}_{13}&=&
1-2\tilde{X}^{ee}_3,
\label{theta13tilde}\\
\cos2\tilde{\theta}_{23}&=&\frac
{|\tilde{X}^{\tau e}_3|^2-|\tilde{X}^{\mu e}_3|^2}
{|\tilde{X}^{\tau e}_3|^2+|\tilde{X}^{\mu e}_3|^2},
\label{theta23tilde}\\
\tilde{\delta}&=&-(\varphi_0+\varphi_3+\varphi_9-\varphi'_9)
=-\arg\det\tilde{U}_0
+\arg\tilde{X}^{\mu e}_1
+\arg\tilde{X}^{\mu e}_3
+\arg\tilde{X}^{\tau e}_3.
\label{deltatilde}
\end{eqnarray}
Eqs.\,(\ref{theta12tilde})--(\ref{deltatilde})
are one of the new results of the present paper.
Notice that the quantities $\tilde{X}^{\alpha\beta}_j$
and $\det\tilde{U}_0$ in
Eqs.\,(\ref{theta12tilde})--(\ref{deltatilde})
are expressed in closed form by the known variables
as is seen in Eq.\,(\ref{solx}) on the assumption
that analytical expressions for all the eigenvalues are known.

A remark is in order.
The standard parametrization (\ref{param1}) is not the only
one for $3\times 3$ unitary matrices,
and other parametrizations are possible
as is described in Appendix \ref{appendixb}.
In the three flavor case, there can be at most two 
level-crossings.  Depending on which pair
of the energy eigenvalues gets close at each
level-crossing, the relevant
effective mixing angle varies.
The appropriate parametrization is
the one in which the orthogonal
matrix, which mixes the two energy
eigenstates, is located on the most right-hand side
of the unitary matrix $U$, because
in such a parametrization
the diagonalized matrix looks like
$\cdots O(\tilde{\theta}_{jk})
\mbox{\rm diag}(\cdots,\tilde{E}_j,\cdots,\tilde{E}_k,\cdots)
O(\tilde{\theta}_{jk})^T\cdots$,
and it becomes clear that $\tilde{\theta}_{jk}$
in the orthogonal matrix $O(\tilde{\theta}_{jk})$
plays a role of the effective mixing angle
which mixes the energy eigenstates with
the energy $\tilde{E}_j$ and $\tilde{E}_k$.
Furthermore in order for the effective mixing angle
$\tilde{\theta}_{jk}$
to be consistent with the two flavor
description, $\tilde{\theta}_{jk}$ should
become maximal at the level-crossing.

\section{Two examples}
We can apply the general discussions in sect.\ref{nonadiabatic}
to concrete examples.
In this section we will discuss two examples.\,\footnote{
These were discussed in Ref.\,\cite{Yasuda:2007jp}
in the adiabatic approximation.}
The first one
is the case with non-standard interactions where the matter potential ${\cal A}$
has the same form as that of the standard case in some basis.
The second example
is the one with magnetic moments in which the energy eigenvalues cannot be
expressed as roots of a quadratic equation.

\subsection{The case with non-standard interactions \label{nonstandard}}

The first example is the oscillation probability
in the presence of new physics in 
propagation~\cite{Wolfenstein:1977ue,Guzzo:1991hi,Roulet:1991sm}.
In this case the mass matrix is given by
\begin{eqnarray}
U{\cal E}U^{-1}
+{\cal A}_{NP}
\label{matrixnp}
\end{eqnarray}
where
\begin{eqnarray}
{\cal A}_{NP}&\equiv&\sqrt{2}G_FN_e
\left(
\begin{array}{ccc}
 1+\epsilon_{ee} & \epsilon_{e\mu} & \epsilon_{e\tau}\\
 \epsilon_{e\mu}^\ast & \epsilon_{\mu\mu} & \epsilon_{\mu\tau}\\
 \epsilon_{e\tau}^\ast & \epsilon_{\mu\tau}^\ast & \epsilon_{\tau\tau}
\end{array}
\right).
\nonumber
\end{eqnarray}
The dimensionless quantities
$\epsilon_{\alpha\beta}$ stand for possible deviation from
the standard matter effect.
It is known~\cite{Davidson:2003ha}
that the constraints on the
parameters $\epsilon_{e\mu}$, $\epsilon_{\mu\mu}$,
$\epsilon_{\mu\tau}$ are strong
($|\epsilon_{\alpha\mu}|\simeq {\cal O}(10^{-2})~(\alpha=e,\mu,\tau$)
while those on the parameters
$\epsilon_{ee}, \epsilon_{e\tau}, \epsilon_{\tau\tau}$
are weak
($|\epsilon_{ee}|,~|\epsilon_{e\tau}|, |\epsilon_{\tau\tau}|
\simeq {\cal O}(1)$.
In Ref.\,\cite{Friedland:2005vy} it was found that large values
($\sim {\cal O}(1)$) of the
parameters $\epsilon_{ee}, \epsilon_{e\tau}, \epsilon_{\tau\tau}$ are
consistent with all the experimental data including those of the
atmospheric neutrino data, provided that one of the
eigenvalues of the matrix (\ref{matrixnp}) at high energy limit
becomes zero, and that such a constraint implies
the relation $\epsilon_{\tau\tau}\simeq|\epsilon_{e\tau}|^2/(1+\epsilon_{ee})$.
For simplicity, therefore, we consider the potential matrix
\begin{eqnarray}
{\cal A}_{NP}&=&A
\left(
\begin{array}{ccc}
 1+\epsilon_{ee} & 0 & \epsilon_{e\tau}\\
 0 & 0 & 0\\
 \epsilon_{e\tau}^\ast & 0 & |\epsilon_{e\tau}|^2/(1+\epsilon_{ee})
\end{array}
\right).
\label{potentialnp}
\end{eqnarray}
Then ${\cal A}_{NP}$ can be diagonalized as
\begin{eqnarray}
{\cal A}_{NP} =　e^{i\gamma'\lambda_9}e^{-i\beta\lambda_5}\,
\mbox{\rm diag}\left(\lambda_{e'},0,0\right)
e^{i\beta\lambda_5}e^{-i\gamma'\lambda_9},
\label{np2}
\end{eqnarray}
where
\begin{eqnarray}
\tan\beta&=&\frac{|\epsilon_{e\tau}|}
{1+\epsilon_{ee}}\nonumber\\
\gamma'&\equiv&\frac{1}{2}\mbox{\rm arg}\,(\epsilon_{e\tau})
\nonumber\\
\lambda_{e'}&=&
\frac{A(1+\epsilon_{ee})}{\cos^2\beta}.
\label{lambdaep}
\end{eqnarray}
As is shown in Appendix \ref{appendixc},
the mass matrix (\ref{matrixnp})
can be written as
\begin{eqnarray}
&{\ }&
U{\cal E}U^{-1}+{\cal A}_{NP}
\nonumber\\
&=&
e^{i\gamma'\lambda_9}e^{-i\beta\lambda_5}
e^{-i\phi_9\lambda_9}
e^{-i\phi_3\lambda_3}
\left[
U''{\cal E}U''^{-1}
+\mbox{\rm diag}\left(\lambda_{e'},0,0\right)
\right]
e^{-i\omega_3\lambda_3}
e^{-i\omega_9\lambda_9}
e^{i\beta\lambda_5}e^{-i\gamma'\lambda_9},
\label{matrixnp2}
\end{eqnarray}
where the phases $\phi_3$, $\phi_9$,
$\omega_3$ and $\omega_9$, which are
defined in Appendix \ref{appendixc},
are introduced to make $U''$
consistent with the standard parametrization
(\ref{param1}).  The mixing angles
$\theta''_{jk}$ and the CP phase $\delta''$
in the standard parametrization of $U''$
are defined by
\begin{eqnarray}
\theta''_{12}&=&
\tan^{-1}\left(|c_\beta e^{-i\gamma'}U_{e2}+s_\beta e^{i\gamma'}U_{\tau2}|
/|c_\beta e^{-i\gamma'}U_{e1}+s_\beta e^{i\gamma'}U_{\tau1}|\right),
\label{thetapp12}\\
\theta''_{13}&=&
\sin^{-1}|c_\beta e^{-i\gamma'}U_{e3}+s_\beta e^{i\gamma'}U_{\tau3}|,
\label{thetapp13}\\
\theta''_{23}&=&
\tan^{-1}\left(U_{\mu3}/
|c_\beta e^{-i\gamma'}U_{\tau3}-s_\beta e^{i\gamma'}U_{e3}|\right),
\label{thetapp23}\\
\delta'' &=& -\text{arg} U''_{e3} 
\text{arg} (c_\beta e^{-i\gamma'}U_{e1}+s_\beta e^{i\gamma'}U_{\tau1})
+\text{arg} (c_\beta e^{-i\gamma'}U_{e2}+s_\beta e^{i\gamma'}U_{\tau2}),
\nonumber\\
&{\ }&
-\text{arg} (c_\beta e^{-i\gamma'}U_{e3}+s_\beta e^{i\gamma'}U_{\tau3})
+\text{arg} (c_\beta e^{-i\gamma'}U_{\tau3}-s_\beta e^{i\gamma'}U_{e3}),
\label{deltapp}
\end{eqnarray}
where $c_\beta\equiv\cos\beta$, $s_\beta\equiv\sin\beta$.
The inside of the square bracket in the
mass matrix (\ref{matrixnp2}) has exactly
the same form as that of the standard case
with replacement $\theta_{jk}\to\theta''_{jk}$,
$\delta\to\delta''$ and $A\to\lambda_{e'}$.
Furthermore, at the two level-crossings specified by
$\Delta E_{31}\cos2\theta''_{13} = \lambda_{e'}$ and
$\Delta E_{21}\cos2\theta''_{12} = (c''_{13})^2\lambda_{e'}$,
$\tilde{\theta}''_{13}$ and
$\tilde{\theta}''_{12}$ become $\pi/4$, respectively.
Therefore, $\tilde{\theta}''_{13}$ and $\tilde{\theta}''_{12}$ can
be regarded as the appropriate mixing angles
to describe the nonadiabatic transition
at the two level-crossings.
Hence we can deduce the jumping factors
at the two level-crossings\,\footnote{
The quantity $P_L$ was given first in
Ref.\,\cite{oai:arXiv.org:hep-ph/0402266}
whose result agrees with ours.}:
\begin{eqnarray}
P_H&=&\exp\left(
-\frac{\pi}{2}\cdot\frac{\Delta E_{31}\sin^22\theta''_{13}}
{\cos2\theta''_{13}|d\log A/dt|_{\mbox{\rm\scriptsize resonance}}}
\right)
\label{phnp}\\
P_L&=&\exp\left(
-\frac{\pi}{2}\cdot\frac{\Delta E_{21}\sin^22\theta''_{12}}
{\cos2\theta''_{12}|d\log A/dt|_{\mbox{\rm\scriptsize resonance}}}
\right)
\label{plnp}
\end{eqnarray}

To estimate the effective mixing matrix elements at the origin $L=0$,
we assume that the matter effect $A$ is much larger than
the energy difference $|\Delta E_{jk}|$.
In this case we can ignore
the term ${\cal E}$ in Eq.\,(\ref{matrixb2}),
and Eq.\,(\ref{np2}) indicates that the mixing matrix
$\tilde{U}$ is given by $e^{i\gamma'\lambda_9}e^{-i\beta\lambda_5}$,
and we get
\begin{eqnarray}
|\tilde{U}_{\alpha j}(0)|^2&=&\left(
\begin{array}{ccc}
c_\beta^2&0&s_\beta^2\cr
0&1&0\cr
s_\beta^2&0&c_\beta^2
\end{array}\right).
\label{u0np}
\end{eqnarray}
From Eqs.\,(\ref{p3}), (\ref{phnp}), (\ref{plnp}) and (\ref{u0np}),
we can obtain the transition probability
$P(\nu_\alpha\to\nu_\beta)$ in the case with
the nonstandard neutrino interaction in propagation.

\subsection{The case with large magnetic moments and a magnetic
field \label{magnetic}}
The second example is the case where
there are three active neutrinos with
magnetic moments and a large magnetic
field\,\footnote{
The possibility that the magnetic moments of neutrinos in
a large magnetic field affect the neutrino flavor transition
caught a lot of attention after this idea was applied to the
solar neutrino deficit in
Refs.\,\cite{Cisneros:1970nq,Okun:1986na,Lim:1987tk,Akhmedov:1988uk}.}.
This is an example where the energy eigenvalues
cannot be expressed as roots of a quadratic equation, and this
case demonstrates the usefulness of the KTY formalism.
Here we assume the magnetic interaction of Majorana type
\begin{eqnarray}
\mu_{\alpha\beta}\bar{\nu}_\alpha\,F_{\lambda\kappa}
\sigma^{\lambda\kappa}\,\nu^c_\beta + h. c.,
\label{majorana}
\end{eqnarray}
and in this case the magnetic moments $\mu_{\alpha\beta}$
are real and anti-symmetric in flavor
indices: $\mu_{\alpha\beta}=-\mu_{\beta\alpha}$.
The hermitian matrix\,\footnote{
See Ref.\,\cite{Grimus:1993fz} for derivation of Eq.\,(\ref{matrixb})
from the Dirac Eq.}
\begin{eqnarray}
{\cal M}\equiv
\left(
\begin{array}{cc}
U{\cal E}U^{-1}&{\cal B}\\
{\cal B}^\dagger&U^\ast{\cal E}(U^\ast)^{-1}
\end{array}\right)
\label{matrixb}
\end{eqnarray}
with
\begin{eqnarray}
{\cal B}_{\alpha\beta}\equiv B\,\mu_{\alpha\beta}
\nonumber
\end{eqnarray}
is the mass matrix
for neutrinos and anti-neutrinos without the matter effect
where neutrinos have the magnetic moments $\mu_{\alpha\beta}$
in the magnetic field $B$.

For simplicity we consider the limit
$\theta_{13}\to0$ and $\Delta m^2_{21}\rightarrow0$,
and we assume that all the CP phases vanish\,\footnote{
In the presence of the magnetic interaction (\ref{majorana})
of Majorana type,
the two CP phases, which are absorbed by
redefinition of the charged lepton fields in the
standard case, cannot be absorbed and therefore
become physical.  Here, however, we assume for simplicity
that these CP phases vanish.}.
Then the matrix (\ref{matrixb}) can be rewritten as
\begin{eqnarray}
{\cal M}
=\frac{1}{2}
\left(\begin{array}{rr}
{\bf 1}&i{\bf 1}\cr
i{\bf 1}&{\bf 1}
\end{array}\right)
\left(\begin{array}{cc}
U{\cal E}U^{-1}+i{\cal B}&0\cr
0&U{\cal E}U^{-1}-i{\cal B}
\end{array}\right)
\left(\begin{array}{rr}
{\bf 1}&-i{\bf 1}\cr
-i{\bf 1}&{\bf 1}
\end{array}\right),
\nonumber
\end{eqnarray}
so the problem of diagonalizing the $6\times6$ matrix (\ref{matrixb})
is reduced to diagonalizing the $3\times3$ matrices
$U{\cal E}U^{-1}\pm i{\cal B}$.
Since we are assuming that all the CP phases vanish,
all the matrix elements $U_{\alpha j}$ and
${\cal B}_{\alpha\beta}=-{\cal B}_{\beta\alpha}$
are real, $U{\cal E}U^{-1}\pm i{\cal B}$ can be diagonalized by
a unitary matrix and its complex conjugate:
\begin{eqnarray}
U{\cal E}U^{-1}+ i{\cal B}&=&
\tilde{U}\tilde{{\cal E}}\tilde{U}^{-1}
\label{matrixb2}\\
U{\cal E}U^{-1}- i{\cal B}&=&
\tilde{U}^\ast\tilde{{\cal E}}(\tilde{U}^\ast)^{-1},
\nonumber
\end{eqnarray}
and the equation for motion is given by
\begin{eqnarray}
&{\ }&\frac{d{\ }}{dt}
\left(\begin{array}{c}
\Psi(t)+i\Psi^c(t)\cr
\Psi(t)-i\Psi^c(t)
\end{array}\right)=
\left(\begin{array}{c}
\tilde{U}(t)\tilde{{\cal E}}(t)\tilde{U}^{-1}(t)
\left\{\Psi(t)+i\Psi^c(t)\right\}
\cr
\tilde{U}^\ast(t)\tilde{{\cal E}}(t)(\tilde{U}^\ast)^{-1}(t)
\left\{\Psi(t)-i\Psi^c(t)\right\}
\end{array}\right).
\label{eqb}
\end{eqnarray}

To evaluate the energy eigenvalues and
the jumping probability,
let us simplify
the matrix (\ref{matrixb2}).
Introducing the notations
\begin{eqnarray}
{\cal B}_{\alpha\beta}&=&B\mu_{\alpha\beta}
\equiv\left(
\begin{array}{ccc}
0&-p_0&-q_0\cr
p_0&0&-r_0\cr
q_0&r_0&0
\end{array}\right),
\nonumber\\
\nonumber\\
e^{-i{\theta}_{23}\lambda_7}\,{\cal B}\,
e^{i{\theta}_{23}\lambda_7}
&=&
\left(
\begin{array}{ccc}
0&-p_0c_{23}+q_0s_{23}&-p_0s_{23}-q_0c_{23}\cr
p_0c_{23}-q_0s_{23}&0&-r_0\cr
p_0s_{23}+q_0c_{23}&r_0&0
\end{array}\right)
\nonumber\\
&\equiv&\left(
\begin{array}{ccc}
0&-p&-q\cr
p&0&-r\cr
q&r&0
\end{array}\right),
\label{matrixb5}
\end{eqnarray}
it is shown in Appendix \ref{appendixd}
that Eq.\,(\ref{matrixb2}) can be rewritten as
\begin{eqnarray}
e^{i{\theta}_{23}\lambda_7}\,
e^{i\omega\lambda_2}\,
\left[\mbox{\rm diag}(0,0,\Delta E_{31})\,
+ \Lambda
e^{i\chi\lambda_5}\,
\lambda_2\,
e^{-i\chi\lambda_5}\,
\right]\,
e^{-i\omega\lambda_2}\,
e^{-i{\theta}_{23}\lambda_7}\,,
\label{matrixb4}
\end{eqnarray}
where $\Lambda$, $\omega$ and $\chi$ are
defined by
\begin{eqnarray}
\Lambda&\equiv&\sqrt{p^2+q^2+r^2},
\label{lambda}\\
\omega&\equiv&\tan^{-1}\frac{r}{q},
\label{omega}\\
\chi&\equiv&\tan^{-1}\frac{\sqrt{q^2+r^2}}{p}.
\label{chi}
\end{eqnarray}
Since the we are mainly interested in the effective
mixing angle which mixes the two energy
eigenstates, the matrices $e^{i{\theta}_{23}\lambda_7}e^{i\omega\lambda_2}$
on the left-hand side and
$e^{-i\omega\lambda_2}e^{-i{\theta}_{23}\lambda_7}$
on the right-hand side of
the square bracket in Eq.\,(\ref{matrixb4}) are irrelevant, so
we discuss the following matrix:
\begin{eqnarray}
{\cal M}&\equiv&\mbox{\rm diag}(0,0,\Delta E_{31})\,
-\frac{\Delta E_{31}}{3}\mbox{\bf 1}
+ \Lambda
e^{i\chi\lambda_5}\,
\lambda_2\,
e^{-i\chi\lambda_5},
\label{m0}
\end{eqnarray}
where a matrix which is proportional to identity
was subtracted for convenience in later calculations
so that the trace of ${\cal M}$ vanishes.

The eigenvalues of the matrix ${\cal M}$
are given in Appendix \ref{appendixe}.
In Fig.~\ref{fig1}
the three eigenvalues $t_j~(j=1,2,3)$ which are normalized by
$2\sqrt{\Delta E_{31}^2/9+\Lambda^2/3}$ are depicted as a
function of $u\equiv 3\Lambda^2/\Delta E_{31}^2$.
If $\chi$ is small, then the two of the three
eigenvalues get close to each other, and
$\chi$ can be regarded as the vacuum mixing angle
near the level-crossing in the present case.
In this example, for a large value of $\Lambda\gg \Delta E_{31}$,
the energy eigenvalues are
$0$ and $\pm \Lambda$,
and we found that there is only one
level-crossing for
$|\Delta E_{31}|\sim \Lambda$,
unlike in the standard three flavor case\,\footnote{
In principle one could say that
the other level-crossing is at
$\Lambda=0$, i.e., in vacuum.
However, in vacuum the jumping factor
$P_C$ vanishes, so that the matrix which
involves $P_L$ in Eq.\,(\ref{p3}) becomes
identity.  Hence there is only only
one level-crossing in practice.}.
So in the following we discuss the contribution
from one level-crossing only.

\begin{figure}[tb]
\begin{center}
\includegraphics[width=10cm]{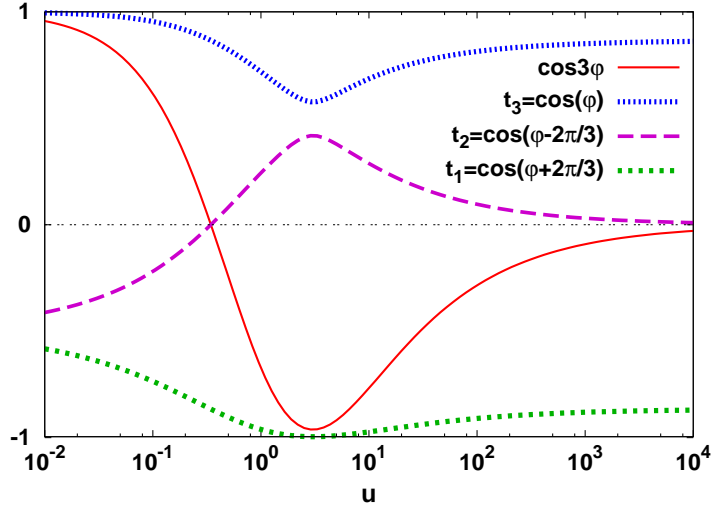}
%
\caption{\label{fig1} 
The behaviors of the normalized eigenvalues
$t_j\equiv\tilde{E}_j/2\sqrt{\Delta E_{31}^2/9+\Lambda^2/3}
=\cos(\varphi+2j\pi/3)\,(j=1,2,3)$ of ${\cal M}$
and $\cos3\varphi$
as functions of
$u\equiv 3\Lambda^2/\Delta E_{31}^2$.
See Appendix \ref{appendixe} for details.}
\end{center}
\end{figure}

Furthermore,
it is shown in Appendix \ref{appendixf}
that the following relation holds:
\begin{eqnarray}
&{\ }&P(\nu_\alpha\rightarrow\nu_\beta)+P(\nu_\alpha\rightarrow\bar{\nu}_\beta)
=\sum_{j,k}
|\tilde{U}_{\beta j}(L)|^2\,
|W_{jk}|^2\,
|\tilde{U}_{\alpha k}^\ast(0)|^2
\nonumber\\
&=&\left(\begin{array}{ccc}
\left|U_{\beta 1}\right|^2&
\left|U_{\beta 2}\right|^2&
\left|U_{\beta 3}\right|^2
\end{array}\right)
\left(\begin{array}{ccc}
1 & 0 &0 \\
0 & 1-P_H & P_H\\
0 & P_H & 1-P_H
\end{array}\right)
\left(\begin{array}{c}
\left|\tilde{U}_{\alpha 1}(0)\right|^2\\
\left|\tilde{U}_{\alpha 2}(0)\right|^2\\
\left|\tilde{U}_{\alpha 3}(0)\right|^2
\end{array}\right),
\label{relationb}
\end{eqnarray}
where we have assumed that the
level-crossing occurs between
the energy eigenstates 2 and 3,
and we have assumed that
there is no magnetic field at the
endpoint $t=L$.

As is explained in detail in Appendix \ref{appendixg},
in the approximation of the small mixing angle $\chi$,
the jumping factor $P_H$ can be calculated as
\begin{eqnarray}
P_H&\simeq& \exp\left( -\frac{\pi}{|d\Lambda/dt|_{u=u_0}}\,\Delta E_{31}^2\,\chi^2
\right).
\label{phb}
\end{eqnarray}

Let us now check whether the exponent
of $P_H$ is proportional to the factor $\gamma$.
For this purpose, we introduce the parametrization of the
mixing matrix other than the standard one:
\begin{eqnarray}
U&=&
e^{i\psi_{12}\lambda_2}\,
\Gamma_\eta^{(13)}\,
e^{i\psi_{13}\lambda_5}\,
(\Gamma_\eta^{(13)})^{-1}\,
e^{i\psi_{23}\lambda_7}
\nonumber\\
&=&
\left(
\begin{array}{ccc}
s_{12}c_{13}&
s_{12}c_{23}-e^{-i\eta}c_{12}s_{13}s_{23}
&s_{12}s_{23}+e^{-i\eta}c_{12}s_{13}c_{23}
\cr
-s_{12}c_{13}&
c_{12}c_{23}+e^{-i\eta}s_{12}s_{13}s_{23}
&c_{12}s_{23}-e^{-i\eta}s_{12}s_{13}c_{23}
\cr
-e^{i\eta}s_{13}&-c_{13}s_{23}&c_{13}c_{23}
\end{array}\right),
\nonumber
\end{eqnarray}
where $c_{jk}\equiv\cos\psi_{jk}$,
$s_{jk}\equiv\sin\psi_{jk}$, and we have
used the notations $\psi_{jk}$ and $\eta$
which are different from those of the
standard parametrization (\ref{param1})
to avoid confusion.
From this expression we observe
\begin{eqnarray}
\tan^2\tilde{\psi}_{23}=
\frac{\tilde{X}_2^{\tau\tau}}
{\tilde{X}_3^{\tau\tau}},
\nonumber
\end{eqnarray}
where $\tilde{X}_j^{\tau\tau}\,(j=2,3)$
can be obtained from Eq.\,(\ref{solx})
by replacing
$U{\cal E}U^{-1}+{\cal A}\to{\cal M}$.
As is shown in Appendix \ref{appendixh},
the effective mixing angle
$\tilde{\psi}_{23}$ becomes maximal at the level-crossing.
So $\tilde{\psi}_{23}$ is the appropriate mixing angle
to describe the nonadiabatic transition
between the two energy eigenstates with
$\tilde{E}_2$ and $\tilde{E}_3$.

In Appendix \ref{appendixh} it is also shown that
the exponent of the
jumping factor $P_H$ coincides with
$-\pi/2$ times the $\gamma$ factor
in the case of a linear potential ($F=1$):
\begin{eqnarray}
\gamma = \left.\frac{\Delta \tilde{E}_{32}}
{2|d\tilde{\psi}_{23}/dt|_{u=u_0}}\right|_{u=u_0}
\simeq\frac{2\Delta E_{31}^2\chi^2}
{|d\Lambda/dt|_{u=u_0}}
=-\frac{\log P_H}{\pi/2}.
\nonumber
\end{eqnarray}

To obtain the transition probability
we also need  the expression for $|\tilde{U}_{\alpha j}(0)|^2$.
We can roughly estimate
the elements $|\tilde{U}_{\alpha j}(0)|^2$
of the effective mixing matrix
at the origin $t=0$ by ignoring
the term ${\cal E}$ in Eq.\,(\ref{matrixb2}).
Using the property (\ref{euler}) we see
\begin{eqnarray}
{\cal B}&=&\left(
\begin{array}{ccc}
0&-p_0&-q_0\cr
p_0&0&-r_0\cr
q_0&r_0&0
\end{array}\right)
=\Lambda\,
e^{i\omega_0\lambda_2}\,
e^{i\chi_0\lambda_5}\,
\lambda_2\,
e^{-i\chi_0\lambda_5}\,
e^{-i\omega_0\lambda_2}\,
\nonumber\\
&=&\Lambda\,
e^{i\omega_0\lambda_2}\,
e^{i\chi_0\lambda_5}\,
e^{i(\pi/4)\lambda_1}\,
\lambda_3\,
e^{-i(\pi/4)\lambda_1}\,
e^{-i\chi_0\lambda_5}\,
e^{-i\omega_0\lambda_2}\,,
\label{euler0}
\end{eqnarray}
where the angles $\omega_0$ and $\chi_0$
are defined by
$\omega_0\equiv\tan^{-1}(r_0/q_0)$,
$\chi_0\equiv\tan^{-1}(\sqrt{q_0^2+r_0^2}/p_0)$.
Eq.\,(\ref{euler0}) implies that the
effective mixing matrix $\tilde{U}$ at the origin is
$\tilde{U}=e^{i\omega_0\lambda_2}e^{i\chi_0\lambda_5}e^{i(\pi/4)\lambda_1}$,
so that we have
\begin{eqnarray}
|\tilde{U}_{\alpha j}(0)|^2&=&\left(
\begin{array}{ccc}
|\frac{1}{\sqrt{2}}(c_{\omega_0}c_{\chi_0}+is_{\omega_0})|^2&
|\frac{1}{\sqrt{2}}(ic_{\omega_0}c_{\chi_0}+s_{\omega_0})|^2&
|c_{\omega_0}s_{\chi_0}|^2\cr
|\frac{1}{\sqrt{2}}(-s_{\omega_0}c_{\chi_0}+ic_{\omega_0})|^2&
|\frac{1}{\sqrt{2}}(-is_{\omega_0}c_{\chi_0}+c_{\omega_0})|^2&
|-s_{\omega_0}s_{\chi_0}|^2\cr
|-\frac{1}{\sqrt{2}}s_{\chi_0}|^2&
|-\frac{i}{\sqrt{2}}s_{\chi_0}|^2&
|c_{\chi_0}|^2
\end{array}\right)
\nonumber\\
&=&\frac{1}{2\Lambda^2(q_0^2+r_0^2)}\,
\left(
\begin{array}{ccc}
r_0^2\Lambda^2+p_0^2q_0^2&
r_0^2\Lambda^2+p_0^2q_0^2&
2q_0^2(q_0^2+r_0^2)\cr
q_0^2\Lambda^2+p_0^2r_0^2&
q_0^2\Lambda^2+p_0^2r_0^2&
2r_0^2(q_0^2+r_0^2)\cr
(q_0^2+r_0^2)^2&
(q_0^2+r_0^2)^2&
2p_0^2(q_0^2+r_0^2)
\end{array}\right).
\label{xb0}
\end{eqnarray}
From Eqs.\,(\ref{relationb}), (\ref{phb}) and
(\ref{xb0}), we obtain the combination
$P(\nu_\alpha\rightarrow\nu_\beta)+P(\nu_\alpha\rightarrow\bar{\nu}_\beta)$
of the transition probabilities.

In the discussions above we have assumed that $|\chi|$ is small.
If this is not the case, then $\tilde{\psi}_{23}$ does not
necessarily become maximal at the
level-crossing.  In that case, instead of
the matrix ${\cal M}$ in Eq.\,(\ref{m0}), we should use
\begin{eqnarray}
{\cal M}'&\equiv&e^{-i\alpha\lambda_5}\,{\cal M}\,
e^{i\alpha\lambda_5}
\equiv
e^{-i\alpha\lambda_5}\,\tilde{U}\tilde{{\cal E}}\tilde{U}^{-1}\,
e^{i\alpha\lambda_5}
\equiv\tilde{U}'\tilde{{\cal E}}(\tilde{U}')^{-1}.
\label{m1}
\end{eqnarray}
$\alpha$ is a parameter which is determined
by the condition
\begin{eqnarray}
\left.\tan^2\tilde{\psi}'_{23}\right|_{u=u_0}=
\left.\frac{\tilde{X}_2^{'\tau\tau}}
{\tilde{X}_3^{'\tau\tau}}\right|_{u=u_0}&=&\left.
\frac{|\tilde{U}_{\tau 2}\cos\alpha+\tilde{U}_{e 2}\sin\alpha|^2}
{|\tilde{U}_{\tau 3}\cos\alpha+\tilde{U}_{e 3}\sin\alpha|^2}
\right|_{u=u_0}=1.
\label{cond1}
\end{eqnarray}
The condition (\ref{cond1}) is a quadratic equation
with respect to $\tan\alpha$:
\begin{eqnarray}
(|\tilde{U}_{e 2}|^2-|\tilde{U}_{e 3}|^2)\,\tan^2\alpha
+2\,\mbox{\rm Re}
(\tilde{U}_{e 2}\tilde{U}_{\tau 2}^\ast
-\tilde{U}_{e 3}\tilde{U}_{\tau 3}^\ast)\,\tan\alpha
+|\tilde{U}_{\tau 2}|^2-|\tilde{U}_{\tau 3}|^2=0,
\label{cond2}
\end{eqnarray}
and the discriminant of Eq.\,(\ref{cond2}) is given by
\begin{eqnarray}
\left\{\mbox{\rm Re}
(\tilde{U}_{e 2}\tilde{U}_{\tau 2}^\ast
-\tilde{U}_{e 3}\tilde{U}_{\tau 3}^\ast)\right\}^2
-(|\tilde{U}_{e 2}|^2-|\tilde{U}_{e 3}|^2)
(|\tilde{U}_{\tau 2}|^2-|\tilde{U}_{\tau 3}|^2).
\label{cond3}
\end{eqnarray}
In the present case, the quantity
$\tilde{X}_j^{e\tau}\equiv\tilde{U}_{e j}\tilde{U}_{\tau j}^\ast
~(j=2,3)$ turns out to be real because
$({\cal M}^{j-1})_{e\tau}~(j=2,3)$ is real.
Hence the discriminant (\ref{cond3}) becomes
$(\tilde{U}_{e 2}\tilde{U}_{\tau 3}
-\tilde{U}_{e 3}\tilde{U}_{\tau 2})^2$
which is positive semi-definite, and the quadratic equation
(\ref{cond2}) always has real roots.
Using one of the roots of (\ref{cond2}) as $\tan\alpha$
in Eq.\,(\ref{m1}), we can evaluate the $\gamma$
factor.  Thus, on the assumption that
the jumping probability $P_H$ is given by the
factor $\exp(-\gamma F \pi/2)$,
we can deduce the jumping factor $P_H$.  In this case,
however, unlike in the case of $|\chi|\ll 1$,
we cannot prove that the exponent
of the jumping probability
(\ref{pc}) in the WKB treatment is equal to
$-\gamma F \pi/2$.

\section{conclusions \label{conclusions}}
Using the formalism which was developed by Kimura, Takamura and
Yokomakura to express analytically the combination
$\tilde{X}^{\alpha\beta}_j\equiv
\tilde{U}_{\alpha j}\tilde{U}_{\beta j\ast}$ of
the mixing matrix elements in matter with constant density,
we have shown that the effective mixing angle
can be analytically expressed in terms of
the mixing matrix elements in vacuum and the energy eigenvalues.
The analytical expression for the effective
mixing angle enables us to evaluate the
nonadiabatic contribution to the transition
probability based on the two assumptions:
(i) The nonadiabatic transitions in the case
with more than two energy eigenstates can
be separately treated as a two state problem
at each level-crossing.
(ii) The exponent of the probability obtained
by the WKB method is proportional to the factor
$\gamma$ which is the ratio of the energy
difference of the two eigenstates to the derivative
of the effective mixing angle at the level-crossing.
We have given two examples: one with
flavor dependent nonstandard interactions
in neutrino propagation and the other with
magnetic moments in a large magnetic field.
In the second example the energy eigenvalues
cannot be expressed as roots of a quadratic
equation and discussions become much less
trivial compared with the standard case
or with the first example.
In the second example we have shown 
in the approximation of the small
mixing angle that the above assumption (ii)
is correct.  If the two
assumptions (i) and (ii) above are correct, then
the KTY formalism enables us to express the
probability of nonadiabatic transitions
in terms of the mixing matrix elements in vacuum
and the energy eigenvalues in general cases.

\appendix
\section{
The effective mixing angle in the standard
three flavor case
\label{appendixa}}
Near the level-crossing
$\Delta E_{31}\simeq A$,
in the leading order in $\Delta E_{21}/\Delta E_{31}$,
the mass matrix (\ref{sch3}) becomes
\begin{eqnarray}
&{\ }&U{\cal E}U^{-1}+{\cal A}
\nonumber\\
& =& 
e^{i\theta_{23}\lambda_7}\,\Gamma_\delta^{(13)}\,
e^{i\tilde{\theta}_{13}\lambda_5}\,\frac{1}{2}
\mbox{\rm diag}\left(\Delta E_{31}+A-\Delta\tilde{E}_{31},
0,\Delta E_{31}+A+\Delta\tilde{E}_{31}\right)\,
\nonumber\\
&{\ }&\times
e^{-i\tilde{\theta}_{13}\lambda_5}\,
(\Gamma_\delta^{(13)})^{-1}\,
e^{-i\theta_{23}\lambda_7},
\nonumber
\end{eqnarray}
where $\lambda_j~(j=5,7)$ are defined by
Eqs.\,(\ref{lambda5}) and (\ref{lambda7}), and
\begin{eqnarray}
\tan2\tilde{\theta}_{13}&\equiv&
\frac{\Delta E_{31}\sin2\theta_{13}}
{\Delta E_{31}\cos2\theta_{13}-A}
\nonumber\\
\Delta\tilde{E}_{31}&\equiv&
\sqrt{(\Delta E_{31}\cos2\theta_{13}-A)^2
+(\Delta E_{31}\sin2\theta_{13})^2}
\nonumber\\
\Gamma_\delta^{(13)}&\equiv&\mbox{\rm diag}(e^{-i\delta/2},1,e^{i\delta/2})
\nonumber
\end{eqnarray}
At $\Delta E_{31}\cos2\theta_{13} = A$,
the effective mixing angle
$\tilde{\theta}_{13}$ becomes $\pi/4$, as in the
two flavor case.  Hence $\tilde{\theta}_{13}$ can
be regarded as the appropriate mixing angle
to describe the nonadiabatic transition
between the two energy eigenstates with
$\tilde{E}_1$ and $\tilde{E}_3$.

On the other hand, near the level-crossing
$\Delta E_{21}\simeq A$,
to first order in $\Delta E_{21}/\Delta E_{31}$
the mass matrix (\ref{sch3}) becomes
\begin{eqnarray}
&{\ }&U{\cal E}U^{-1}+{\cal A}
\nonumber\\
&=& 
e^{i\theta_{23}\lambda_7}\,
\Gamma_\delta^{(13)}\,
e^{i\theta_{13}\lambda_5}\,
(\Gamma_\delta^{(13)})^{-1}\,
\exp\left(i\sum_{j=4}^7d_j\lambda_j\right)
e^{i\tilde{\theta}_{12}\lambda_2}
\nonumber\\
&{\ }&\times
\left[\mbox{\rm diag}\left\{\frac{1}{2}
(\Delta E_{21}+Ac^2_{13}+\Delta\tilde{E}_{21},
\Delta E_{21}+Ac^2_{13}-\Delta\tilde{E}_{21},
\Delta E_{31}+As_{13}^2\right\}
\right]
\nonumber\\
&{\ }&\times
e^{-i\tilde{\theta}_{12}\lambda_2}
\exp\left(-i\sum_{j=4}^7d_j\lambda_j\right)
\Gamma_\delta^{(13)}\,
e^{-i\theta_{13}\lambda_5}\,
(\Gamma_\delta^{(13)})^{-1}\,
e^{-i\theta_{23}\lambda_7},
\nonumber
\end{eqnarray}
where $\lambda_2$ is defined by
Eq.\,(\ref{lambda2}),
\begin{eqnarray}
\tan2\tilde{\theta}_{12}&\equiv&
\frac{\Delta E_{21}\sin2\theta_{12}}
{\Delta E_{21}\cos2\theta_{12}-Ac^2_{13}}
\nonumber\\
\Delta\tilde{E}_{21}&\equiv&
\sqrt{(\Delta E_{21}\cos2\theta_{12}-Ac^2_{13})^2
+(\Delta E_{21}\sin2\theta_{12})^2}
\end{eqnarray}
and $d_j~(j=4,\cdots,7)$ are the small coefficients 
given by
$d_4=c_{12}\sin\delta A/\Delta E_{31}$,
$d_5=c_{12}\cos\delta A/\Delta E_{31}$,
$d_6=s_{12}\sin\delta A/\Delta E_{31}$
and
$d_7=s_{12}\cos\delta A/\Delta E_{31}$.
At $\Delta E_{21}\cos2\theta_{12} = Ac_{13}^2$,
the effective mixing angle
$\tilde{\theta}_{12}$ becomes $\pi/4$.
Hence $\tilde{\theta}_{12}$ can
be regarded as the appropriate mixing angle
to describe the nonadiabatic transition
between the two energy eigenstates with
$\tilde{E}_1$ and $\tilde{E}_2$.

\section{
Parametrizations for $3\times 3$ unitary matrices
\label{appendixb}}
Assuming the same form as for the standard parametrization,
we can consider six permutations for $3\times 3$ unitary matrices
as follows:
\begin{eqnarray}
U&=&
e^{i\theta_{23}\lambda_7}
\,\Gamma_\delta^{(13)}\,
e^{i\theta_{13}\lambda_5}
\,(\Gamma_\delta^{(13)})^{-1}\,
e^{i\theta_{12}\lambda_2}
\nonumber\\
&=&
\left(
\begin{array}{ccc}
1 & 0 & 0\cr
0 & c_{23} & s_{23}\cr
0 & -s_{23} & c_{23}
\end{array}\right)
\left(
\begin{array}{ccc}
e^{-i\delta/2} & 0 & 0\cr
0 & 1 & 0\cr
0 & 0 & e^{i\delta/2}
\end{array}\right)
\left(
\begin{array}{ccc}
c_{13} & 0 &  s_{13}\cr
0 & 1 & 0\cr
-s_{13}& 0 & c_{13}
\end{array}\right)
\nonumber\\
&{\ }&\times
\left(
\begin{array}{ccc}
e^{i\delta/2} & 0 & 0\cr
0 & 1 & 0\cr
0 & 0 & e^{-i\delta/2}
\end{array}\right)
\left(
\begin{array}{ccc}
c_{12} & s_{12} & 0\cr
-s_{12} & c_{12} & 0\cr
0 & 0 & 1
\end{array}\right),
\nonumber\\
&=&
\left(
\begin{array}{ccc}
c_{12}c_{13} & s_{12}c_{13} &  s_{13}e^{-i\delta}
\cr
-s_{12}c_{23}-c_{12}s_{23}s_{13}e^{i\delta} & 
c_{12}c_{23}-s_{12}s_{23}s_{13}e^{i\delta} & 
s_{23}c_{13}
\cr
s_{12}s_{23}-c_{12}c_{23}s_{13}e^{i\delta} & 
-c_{12}s_{23}-s_{12}c_{23}s_{13}e^{i\delta} & 
c_{23}c_{13}
\end{array}\right),
\label{param1}
\end{eqnarray}

\begin{eqnarray}
U&=&
e^{i\theta_{13}\lambda_5}
\,\Gamma_\delta^{(23)}\,
e^{i\theta_{23}\lambda_7}
\,(\Gamma_\delta^{(23)})^{-1}\,
e^{i\theta_{12}\lambda_2}
\nonumber\\
&=&
\left(
\begin{array}{ccc}
c_{12}c_{13}+e^{i\delta}s_{12}s_{13}s_{23}&
s_{12}c_{13}-e^{i\delta}c_{12}s_{13}s_{23}&
s_{13}c_{23}
\cr
-s_{12}c_{23}&
c_{12}c_{23}&
s_{23}e^{-i\delta}
\cr
c_{12}s_{13}+e^{i\delta}s_{12}c_{13}s_{23}&
-s_{12}s_{13}-e^{i\delta}c_{12}c_{13}s_{23}&
c_{13}c_{23}
\end{array}\right),
\label{param2}
\end{eqnarray}

\begin{eqnarray}
U&=&
e^{i\theta_{23}\lambda_7}
\,\Gamma_\delta^{(12)}\,
e^{i\theta_{12}\lambda_2}
\,(\Gamma_\delta^{(12)})^{-1}\,
e^{i\theta_{13}\lambda_5}
\nonumber\\
&=&
\left(
\begin{array}{ccc}
c_{12}c_{13}&
e^{-i\delta}s_{12}&
c_{12}s_{13}
\cr
-s_{13}s_{23}-e^{i\delta}s_{12}c_{13}c_{23}&
c_{12}c_{23}&
c_{13}s_{23}-e^{i\delta}s_{12}s_{13}c_{23}
\cr
-s_{13}c_{23}-e^{i\delta}s_{12}c_{13}s_{23}&
-c_{12}s_{23}&
c_{13}c_{23}+e^{i\delta}s_{12}s_{13}s_{23}
\end{array}\right),
\label{param3}
\end{eqnarray}

\begin{eqnarray}
U&=&
e^{i\theta_{12}\lambda_2}
\,\Gamma_\delta^{(23)}\,
e^{i\theta_{23}\lambda_7}
\,(\Gamma_\delta^{(23)})^{-1}\,
e^{i\theta_{13}\lambda_5}
\nonumber\\
&=&
\left(
\begin{array}{ccc}
c_{12}c_{13}-e^{-i\delta}s_{12}s_{13}s_{23}&
s_{12}c_{23}&
c_{12}s_{13}-e^{-i\delta}s_{12}c_{13}s_{23}
\cr
-s_{12}c_{13}-e^{-i\delta}c_{12}s_{13}s_{23}&
c_{12}c_{23}&
-s_{12}s_{13}+e^{-i\delta}c_{12}c_{13}s_{23}
\cr
-s_{13}c_{23}&
-e^{i\delta}s_{23}&
c_{13}c_{23}
\end{array}\right),
\label{param4}
\end{eqnarray}

\begin{eqnarray}
U&=&
e^{i\theta_{12}\lambda_2}
\,\Gamma_\delta^{(13)}\,
e^{i\theta_{13}\lambda_5}
\,(\Gamma_\delta^{(13)})^{-1}\,
e^{i\theta_{23}\lambda_7}
\nonumber\\
&=&
\left(
\begin{array}{ccc}
c_{12}c_{13}&
s_{12}c_{23}-e^{-i\delta}c_{12}s_{13}s_{23}&
s_{12}s_{23}+e^{-i\delta}c_{12}s_{13}s_{23}
\cr
-s_{12}c_{13}&
c_{12}c_{23}+e^{-i\delta}s_{12}s_{13}s_{23}&
c_{12}s_{23}-e^{-i\delta}s_{12}s_{13}c_{23}
\cr
-e^{i\delta}s_{13}&
-c_{13}s_{23}&
c_{13}c_{23}
\end{array}\right),
\label{param5}
\end{eqnarray}

\begin{eqnarray}
U&=&
e^{i\theta_{13}\lambda_2}
\,\Gamma_\delta^{(12)}\,
e^{i\theta_{12}\lambda_5}
\,(\Gamma_\delta^{(12)})^{-1}\,
e^{i\theta_{23}\lambda_7}
\nonumber\\
&=&
\left(
\begin{array}{ccc}
c_{12}c_{13}&
-s_{13}c_{23}+e^{-i\delta}s_{12}c_{13}c_{23}&
s_{13}c_{23}+e^{-i\delta}s_{12}c_{13}s_{23}
\cr
-e^{i\delta}s_{12}&
c_{12}c_{23}&
c_{12}s_{23}
\cr
-c_{12}s_{13}&
-c_{13}s_{23}-e^{-i\delta}s_{12}s_{13}c_{23}&
c_{13}c_{23}-e^{-i\delta}s_{12}s_{13}s_{23}
\end{array}\right).
\label{param6}
\end{eqnarray}
Here we have introduced the notation:
\begin{eqnarray}
\Gamma_\delta^{(12)}&\equiv&
\mbox{\rm diag}(e^{-i\delta/2},e^{i\delta/2},1),
\nonumber\\
\Gamma_\delta^{(13)}&\equiv&
\mbox{\rm diag}(e^{-i\delta/2},1,e^{i\delta/2}),
\nonumber\\
\Gamma_\delta^{(23)}&\equiv&
\mbox{\rm diag}(1,e^{-i\delta/2},e^{i\delta/2}).
\nonumber
\end{eqnarray}

\section{
The derivation of Eq.\,(\ref{matrixnp2})
\label{appendixc}}
Because of the form of the matrix ${\cal A}_{NP}$
(\ref{np2}),
the mass matrix (\ref{matrixnp}) can be
written as\,\cite{Yasuda:2007jp}
\begin{eqnarray}
&{\ }&
U{\cal E}U^{-1}+{\cal A}_{NP}
\nonumber\\
&=&
e^{i\gamma'\lambda_9}e^{-i\beta\lambda_5}
\left[
e^{i\beta\lambda_5}e^{-i\gamma'\lambda_9}U{\cal E}U^{-1}
e^{i\gamma'\lambda_9}e^{-i\beta\lambda_5}
+\mbox{\rm diag}\left(\lambda_{e'},0,0\right)
\right]
e^{i\beta\lambda_5}e^{-i\gamma'\lambda_9}.
\label{massmnp}
\end{eqnarray}
Here we introduce the following two unitary matrices:
\begin{eqnarray}
U'&\equiv& e^{i\beta\lambda_5}e^{-i\gamma'\lambda_9}\,U
\nonumber\\
&=&\left(\begin{array}{ccc}
c_\beta e^{-i\gamma'}U_{e1}+s_\beta e^{i\gamma'}U_{\tau1}&
c_\beta e^{-i\gamma'}U_{e2}+s_\beta e^{i\gamma'}U_{\tau2}&
c_\beta e^{-i\gamma'}U_{e3}+s_\beta e^{i\gamma'}U_{\tau3}\cr
U_{\mu1}&U_{\mu2}&U_{\mu3}\cr
c_\beta e^{-i\gamma'}U_{\tau1}-s_\beta e^{i\gamma'}U_{e1}&
c_\beta e^{-i\gamma'}U_{\tau2}-s_\beta e^{i\gamma'}U_{e2}&
c_\beta e^{-i\gamma'}U_{\tau3}-s_\beta e^{i\gamma'}U_{e3}
\end{array}
\right),
\label{uprime}\\
U''&\equiv&e^{i\phi_3\lambda_3}e^{i\phi_9\lambda_9}\,
U'\,e^{i\omega_9\lambda_9}e^{i\omega_3\lambda_3},
\nonumber
\end{eqnarray}
where $c_\beta\equiv\cos\beta$, $s_\beta\equiv\sin\beta$,
and $U''$ is in the standard parametrization (\ref{param1}).
The phases $\phi_3$, $\phi_9$,
$\omega_3$ and $\omega_9$ are defined in such a way
that the elements $U''_{e1}$, $U''_{e2}$, $U''_{\mu 3}$, $U''_{\tau3}$
become real to be consistent with the standard parametrization
(\ref{param1}), and are given by
\begin{eqnarray}
\phi_3&=&-\frac{1}{3}\text{arg} U'_{e1}
-\frac{1}{3}\text{arg} U'_{e2}
-\frac{2}{3}\text{arg} U'_{\tau 3},
\nonumber\\
\phi_9&=&-\frac{1}{3}\text{arg} U'_{e1}
-\frac{1}{3}\text{arg} U'_{e2}
+\frac{1}{3}\text{arg} U'_{\tau 3},
\nonumber\\
\omega_3&=&-\frac{2}{3}\text{arg} U'_{e1}
+\frac{1}{3}\text{arg} U'_{e2}
-\frac{2}{3}\text{arg} U'_{\tau 3},
\nonumber\\
\omega_9&=&\frac{1}{3}\text{arg} U'_{e1}
+\frac{1}{3}\text{arg} U'_{e2}
+\frac{2}{3}\text{arg} U'_{\tau 3}.
\nonumber
\end{eqnarray}
In these expressions, $\text{arg} U'_{\alpha j}$
can be read off from Eq.\,(\ref{uprime}).
Thus we obtain the
expression for the three mixing angles $\theta''_{jk}$
(\ref{thetapp12}), (\ref{thetapp13}), (\ref{thetapp23})
and the Dirac phase $\delta''$
(\ref{deltapp})
in $U''$\,\footnote{
There was an error in the discussion on the phases in Appendix C
in Ref.\,\cite{Yasuda:2007jp}
in the case of the non-standard interaction.  The expressions
of the phases here correct those in Ref.\,\cite{Yasuda:2007jp}.}:
\begin{eqnarray}
\theta''_{12}&=&\tan^{-1}(U''_{e2}/U''_{e1})=
\tan^{-1}\left(|c_\beta e^{-i\gamma'}U_{e2}+s_\beta e^{i\gamma'}U_{\tau2}|
/|c_\beta e^{-i\gamma'}U_{e1}+s_\beta e^{i\gamma'}U_{\tau1}|\right),
\nonumber\\
\theta''_{13}&=&\sin^{-1}|U''_{e3}|=
\sin^{-1}|c_\beta e^{-i\gamma'}U_{e3}+s_\beta e^{i\gamma'}U_{\tau3}|,
\nonumber\\
\theta''_{23}&=&\tan^{-1}(U''_{\mu3}/U''_{\tau3})=
\tan^{-1}\left(U_{\mu3}/
|c_\beta e^{-i\gamma'}U_{\tau3}-s_\beta e^{i\gamma'}U_{e3}|\right),
\nonumber\\
\delta'' &=& -\text{arg} U''_{e3} 
\nonumber\\
&=&
\text{arg} (c_\beta e^{-i\gamma'}U_{e1}+s_\beta e^{i\gamma'}U_{\tau1})
+\text{arg} (c_\beta e^{-i\gamma'}U_{e2}+s_\beta e^{i\gamma'}U_{\tau2})
\nonumber\\
&{\ }&
-\text{arg} (c_\beta e^{-i\gamma'}U_{e3}+s_\beta e^{i\gamma'}U_{\tau3})
+\text{arg} (c_\beta e^{-i\gamma'}U_{\tau3}-s_\beta e^{i\gamma'}U_{e3}).
\nonumber
\end{eqnarray}

\section{
The derivation of Eq.  (\ref{matrixb4})
\label{appendixd}}
Because of the form of the matrix
(\ref{matrixb5}), we have
\begin{eqnarray}
U{\cal E}U^{-1}+ i{\cal B}&=&
e^{i{\theta}_{23}\lambda_7}\,
\mbox{\rm diag}(0,0,\Delta E_{31})\,
e^{-i{\theta}_{23}\lambda_7}\,
+ i\left(
\begin{array}{ccc}
0&-p_0&-q_0\cr
p_0&0&-r_0\cr
q_0&r_0&0
\end{array}\right)
\nonumber\\
&=&
e^{i{\theta}_{23}\lambda_7}\,
\left[\mbox{\rm diag}(0,0,\Delta E_{31})\,
+ i
\left(
\begin{array}{ccc}
0&-p&-q\cr
p&0&-r\cr
q&r&0
\end{array}\right)
\right]
e^{-i{\theta}_{23}\lambda_7}\,.
\label{matrixb3}
\end{eqnarray}
$T_1=i\lambda_7$, $T_2=-i\lambda_5$, $T_3=i\lambda_2$
are the generators of the $SO(3)$ group, and
the real anti-symmetric matrix 
$e^{i{\theta}_{23}\lambda_7}\,{\cal B}\,
e^{-i{\theta}_{23}\lambda_7}
=-i(p\lambda_2+q\lambda_5+r\lambda_7)$
can be rewritten as
\begin{eqnarray}
-i(p\lambda_2+q\lambda_5+r\lambda_7)
&=&-i
e^{i\omega\lambda_2}\,
\,\left(
p\lambda_2+
\sqrt{q^2+r^2}\,
\lambda_7
\right)\,
e^{-i\omega\lambda_2}
\nonumber\\
&=&-i
e^{i\omega\lambda_2}\,
e^{i\chi\lambda_5}\,
\Lambda\lambda_2\,
e^{-i\chi\lambda_5}\,
e^{-i\omega\lambda_2}\,,
\label{euler}
\end{eqnarray}
where $\Lambda$, $\omega$ and $\chi$ are defined
in Eqs.\,(\ref{lambda}), (\ref{omega}) and (\ref{chi}).
Hence the matrix (\ref{matrixb3}) becomes
\begin{eqnarray}
U{\cal E}U^{-1}+ i{\cal B}&=&
e^{i{\theta}_{23}\lambda_7}\,
\left[\mbox{\rm diag}(0,0,\Delta E_{31})\,
+ \Lambda
e^{i\omega\lambda_2}\,
e^{i\chi\lambda_5}\,
\lambda_2\,
e^{-i\chi\lambda_5}\,
e^{-i\omega\lambda_2}
\right]
e^{-i{\theta}_{23}\lambda_7}
\nonumber\\
&=&e^{i{\theta}_{23}\lambda_7}\,
e^{i\omega\lambda_2}\,
\left[\mbox{\rm diag}(0,0,\Delta E_{31})\,
+ \Lambda
e^{i\chi\lambda_5}\,
\lambda_2\,
e^{-i\chi\lambda_5}\,
\right]\,
e^{-i\omega\lambda_2}\,
e^{-i{\theta}_{23}\lambda_7}.
\nonumber
\end{eqnarray}

\section{
The energy eigenvalues
in the case with large magnetic moments and a magnetic
field
\label{appendixe}}
The eigenvalue of the matrix in Eq.\,(\ref{m0})
can be obtained from
the eigenvalue equation
\begin{eqnarray}
0&=&|\tilde{E}\,\mbox{\bf 1}-{\cal M}|
=\tilde{E}^3-
\left(\Lambda^2+\frac{\Delta E_{31}^2}{3}
\right)\tilde{E}
-\frac{2}{27}
\Delta E_{31}^3
+\frac{1+3\cos2\chi}{6}\,\Lambda^2\Delta E_{31}.
\label{cubiceq}
\end{eqnarray}
The three roots of the cubic equation (\ref{cubiceq}) are given by
\begin{eqnarray}
\tilde{E}_1&=&2R\cos(\varphi+\frac{2}{3}\pi),~
\nonumber\\
\tilde{E}_2&=&2R\cos(\varphi-\frac{2}{3}\pi),~
\nonumber\\
\tilde{E}_3&=&2R\cos\varphi,
\label{tj}
\end{eqnarray}
where
\begin{eqnarray}
R&\equiv& 
\left(
\frac{\Delta E_{31}^2}{9}
+\frac{\Lambda^2}{3}
\right)^{1/2},\nonumber\\
\cos3\varphi&\equiv&
\frac{1}{R^3}\,
\left\{\left(\frac{\Delta E_{31}}{3}\right)^3
-\frac{1+3\cos2\chi}{12}\,\Lambda^2\Delta E_{31}
\right\}
 = \frac{1-Du}{(1+u)^{3/2}},
\label{cos3}\\
u&\equiv& \frac{3\Lambda^2}{\Delta E_{31}},
\nonumber\\
D&\equiv&\frac{3}{4}(1+3\cos2\chi).
\nonumber
\end{eqnarray}
The extremum of $\cos3\varphi$ is given by
the condition
\begin{eqnarray}
0 = \frac{d{\ }}{du}\cos3\varphi
 = \frac{d{\ }}{du}
\left\{\frac{1-Du}{(1+u)^{3/2}}\right\}
 = \frac{D(u-2-3/D)}{2(1+u)^{5/2}},
\nonumber
\end{eqnarray}
so
\begin{eqnarray}
u = u_0 \equiv 2+\frac{3}{D}
\label{lv}
\end{eqnarray}
gives the condition for the level-crossing.
At $u = u_0$, the value of $\cos3\varphi$ is
\begin{eqnarray}
\left.\cos3\varphi\right|_{u=u_0}
= \left.\frac{1-Du}{(1+u)^{3/2}}\right|_{u=u_0}
= -\left(\frac{D}{3}\right)^{\frac{3}{2}}
\sqrt{\frac{4}{1+D}}
\nonumber
\end{eqnarray}

When $|\chi|$ is small, we have
\begin{eqnarray}
D&\simeq& 3-\frac{9}{2}\chi^2,
\label{dsmall}\\
u_0&\simeq& 3 + \frac{3}{2} \chi^2,
\label{u0small}
\end{eqnarray}
so that
$\displaystyle\left.\cos3\varphi\right|_{u=u_0}$
approaches -1:
\begin{eqnarray}
\displaystyle\left.\cos3\varphi\right|_{u=u_0}
\simeq -1 + \frac{27}{16}\chi^2
\nonumber
\end{eqnarray}
This implies that $\left.\varphi\right|_{u=u_0}$
is close to $\pi/3$:
\begin{eqnarray}
\displaystyle\left.\varphi\right|_{u=u_0}
\simeq \frac{\pi}{3} - \sqrt{\frac{3}{8}}\chi
\label{phismall}
\end{eqnarray}
At $u=0$, we have $\cos3\varphi=1$ which
implies $\left.\varphi\right|_{u=u_0}=0$.
As $u$ varies from 0 to $u_0$, therefore,
$\varphi$ varies from $0$ to $\pi/3 - \sqrt{3/8}\,\chi$.

To see the behaviors of the level-crossing,
it is useful to plot 
the normalized eigenvalues
$t_j\equiv\tilde{E}_j/2R\,(j=1,2,3)$ of ${\cal M}$,
instead of $\tilde{E}_j$ themselves.
The values of $t_j\,(j=1,2,3)$ are
shown together with $\cos3\varphi$
in Fig.~\ref{fig1} for $D=2.9$
as functions of
$u\equiv 3\Lambda^2/\Delta E_{31}^2$.

\section{
The derivation of the transition probability
in the case with large magnetic moments and a magnetic
field
\label{appendixf}}
As is explained in the main text, 
we discuss the contribution
from one level-crossing only.
By taking into account the nonadiabatic contribution,
we can integrate Eq.\,(\ref{eqb}):
\begin{eqnarray}
\Psi(L)+i\Psi^c(L)&=&
\tilde{U}(L)\,e^{-i\Phi_2}\,W\,e^{-i\Phi_1}\,\tilde{U}^{-1}(0)
\left(\Psi(0)+i\Psi^c(0)\right)
\nonumber\\
\Psi(L)-i\Psi^c(L)&=&
\tilde{U}^\ast(L)\,e^{-i\Phi_2}\,W^\ast\,e^{-i\Phi_1}\,(\tilde{U}^\ast)^{-1}(0)
\left(\Psi(0)-i\Psi^c(0)\right)
\nonumber
\end{eqnarray}
where $W$ stands for the transition matrix
between the two energy eigenstates at the
level-crossing $t=t_R$, and
\begin{eqnarray}
\Phi_1&\equiv& \int_0^{t_R}\tilde{{\cal E}}(t)\,dt,
\nonumber\\
\Phi_2&\equiv& \int_{t_R}^L\tilde{{\cal E}}(t)\,dt.
\nonumber
\end{eqnarray}
From this we get
\begin{eqnarray}
\Psi(L)&=&\frac{1}{2}\left[
\tilde{U}(L)\,e^{-i\Phi_2}\,W\,e^{-i\Phi_1}\,\tilde{U}^{-1}(0)
\left(\Psi(0)+i\Psi^c(0)\right)\right.
\nonumber\\
&{\ }&\left.
+\tilde{U}^\ast(L)\,e^{-i\Phi_2}\,W^\ast\,e^{-i\Phi_1}\,(\tilde{U}^\ast)^{-1}(0)
\left(\Psi(0)-i\Psi^c(0)\right)
\right]
\nonumber
\end{eqnarray}
Thus the probability amplitudes
for the flavor transition are given by
\begin{eqnarray}
A(\nu_\alpha\rightarrow\nu_\beta)
&=&
\frac{1}{2}\left[
\tilde{U}(L)\,e^{-i\Phi_2}\,W\,e^{-i\Phi_1}\,\tilde{U}^{-1}(0)
+\tilde{U}^\ast(L)\,e^{-i\Phi_2}\,W^\ast\,e^{-i\Phi_1}\,(\tilde{U}^\ast)^{-1}(0)
\right]_{\beta\alpha}
\nonumber\\
&=&\sum_{j,k}\left(e^{-i\Phi_2}\right)_{jj}
\left(e^{-i\Phi_1}\right)_{kk}
\mbox{\rm Re}\left\{\tilde{U}_{\beta j}(L)\,
W_{jk}\,
\tilde{U}_{\alpha k}^\ast(0)
\right\}
\nonumber\\
A(\nu_\alpha\rightarrow\bar{\nu}_\beta)
&=&
\frac{i}{2}\left[
\tilde{U}(L)\,e^{-i\Phi_2}\,W\,e^{-i\Phi_1}\,\tilde{U}^{-1}(0)
-\tilde{U}^\ast(L)\,e^{-i\Phi_2}\,W^\ast\,e^{-i\Phi_1}\,(\tilde{U}^\ast)^{-1}(0)
\right]_{\beta\alpha}
\nonumber\\
&=&-\sum_{j,k}\left(e^{-i\Phi_2}\right)_{jj}
\left(e^{-i\Phi_1}\right)_{kk}
\mbox{\rm Im}\left\{\tilde{U}_{\beta j}(L)\,
W_{jk}\,
\tilde{U}_{\alpha k}^\ast(0)
\right\}
\nonumber
\end{eqnarray}
Hence we obtain the transition probabilities
\begin{eqnarray}
P(\nu_\alpha\rightarrow\nu_\beta)
&=&\sum_{j,k,j',k'}
\left(e^{-i\Phi_2}\right)_{jj}
\left(e^{i\Phi_2}\right)_{j'j'}
\left(e^{-i\Phi_1}\right)_{kk}
\left(e^{i\Phi_1}\right)_{k'k'}
\nonumber\\
&{\ }&\times
\mbox{\rm Re}\left\{\tilde{U}_{\beta j}(L)\,
W_{jk}\,
\tilde{U}_{\alpha k}^\ast(0)
\right\}
\mbox{\rm Re}\left\{\tilde{U}_{\beta j'}(L)\,
W_{j'k'}\,
\tilde{U}_{\alpha k'}^\ast(0)
\right\}
\nonumber\\
&\to&\sum_{j,k}
\left[
\mbox{\rm Re}\left\{\tilde{U}_{\beta j}(L)\,
W_{jk}\,
\tilde{U}_{\alpha k}^\ast(0)
\right\}
\right]^2
\nonumber\\
P(\nu_\alpha\rightarrow\bar{\nu}_\beta)
&\to&\sum_{j,k}
\left[
\mbox{\rm Im}\left\{\tilde{U}_{\beta j}(L)\,
W_{jk}\,
\tilde{U}_{\alpha k}^\ast(0)
\right\}
\right]^2,
\label{probb}
\end{eqnarray}
where we have taken the limit $t_R\to\infty$,
$L\to\infty$ and we have averaged over rapid
oscillations:
$(e^{-i\Phi_1})_{kk}(e^{i\Phi_1})_{k'k'}\to\delta_{kk'}$,
$(e^{-i\Phi_2})_{jj}(e^{i\Phi_2})_{j'j'}\to\delta_{jj'}$.
Each probability in Eqs.\,(\ref{probb}) itself is not expressed in
terms of $\tilde{X}^{\alpha\alpha}_j(0)$, but we find that the
following relation holds:
\begin{eqnarray}
&{\ }&P(\nu_\alpha\rightarrow\nu_\beta)+P(\nu_\alpha\rightarrow\bar{\nu}_\beta)
=\sum_{j,k}
|\tilde{U}_{\beta j}(L)|^2\,
|W_{jk}|^2\,
|\tilde{U}_{\alpha k}^\ast(0)|^2
\nonumber\\
&=&\left(\begin{array}{ccc}
\left|U_{\beta 1}\right|^2&
\left|U_{\beta 2}\right|^2&
\left|U_{\beta 3}\right|^2
\end{array}\right)
\left(\begin{array}{ccc}
1 & 0 &0 \\
0 & 1-P_H & P_H\\
0 & P_H & 1-P_H
\end{array}\right)
\left(\begin{array}{c}
\left|\tilde{U}_{\alpha 1}(0)\right|^2\\
\left|\tilde{U}_{\alpha 2}(0)\right|^2\\
\left|\tilde{U}_{\alpha 3}(0)\right|^2
\end{array}\right),
\nonumber
\end{eqnarray}
where we have assumed that the
level-crossing occurs between
the energy eigenstates 2 and 3,
we have used the fact that
in that case
$|W_{23}|^2=|W_{32}|^2=P_H$,
$|W_{22}|^2=|W_{33}|^2=1-P_H$,
and we have assumed that
there is no magnetic field at the
endpoint $t=L$.

\section{
The calculations of the jumping factor $P_H$
in the case with large magnetic moments and a magnetic
field
\label{appendixg}}
We have seen in Appendix \ref{appendixe} that
$\varphi$ varies from $0$ to $\pi/3 - \sqrt{3/8}\,\chi$,
as $u$ varies from 0 to $u_0$.
To estimate the jumping probability $P_H$ near the
level-crossing $u=u_0$, let us obtain the complex solution
$u$ of the equation
\begin{eqnarray}
\cos3\varphi&=&\frac{1-Du}{(1+u)^{3/2}}=-1
\label{cos3i}
\end{eqnarray}
for $D<3$.  Eq.\,(\ref{cos3i}) gives
\begin{eqnarray}
u\left\{u^2-(D^2-3)u+3+2D
\right\}=0,
\nonumber
\end{eqnarray}
Thus the solutions other than $u=0$ for Eq.\,(\ref{cos3i}) are
\begin{eqnarray}
u = \frac{D^2-3}{2}\,\pm i\sqrt{(3-D)(D+1)^3}
\simeq 3 \pm 6\sqrt{2}\chi \,i,
\label{ui}
\end{eqnarray}
where we have used the condition
(\ref{dsmall}) for small $|\chi|$.
Thus we take the path of the complex integral
for $P_H$ as
\begin{eqnarray}
u &=&  3 + 6\sqrt{2}\chi\xi \,i,
\quad 0\le\xi\le 1\,.
\label{uc}
\end{eqnarray}
When the complex variable lies in the region
specified by Eq.\,(\ref{uc}), we have
\begin{eqnarray}
\cos3\varphi&=&-1+\frac{27}{16}\,\chi^2(1-\xi^2),
\nonumber\\
3\varphi&=&\pi-\sqrt{\frac{27}{8}}\,\chi\sqrt{1-\xi^2}
\nonumber\\
\varphi&=&\frac{\pi}{3}-\sqrt{\frac{3}{8}}\,\chi\sqrt{1-\xi^2}
\label{phi1}
\end{eqnarray}
From Eq.\,(\ref{tj})
the difference $\Delta \tilde{E}_{32}\equiv \tilde{E}_3-\tilde{E}_2$
of the two eigenvalues is given 
\begin{eqnarray}
\Delta \tilde{E}_{32} &=& 2\sqrt{3}R\sin\left(\frac{\pi}{3}-\varphi\right)
\nonumber\\
&=& 2\sqrt{3}\frac{\Delta E_{31}}{3}
\sqrt{1+u}\,
\sin\left(\frac{\pi}{3}-\varphi\right)
\nonumber\\
&\simeq&\sqrt{2}\Delta E_{31}
\,\chi\sqrt{1-\xi^2},
\nonumber
\end{eqnarray}
where we have used Eq.\,(\ref{phi1}) and
the fact that $|\chi|\ll 1$ in the second line.
Thus the jumping probability near the point $u=u_0$
is given by
\begin{eqnarray}
P_H&=& \exp\left[ -\,\mbox{\rm Im}\left(
\int_{\xi=0}^{\xi=1}\,
\Delta \tilde{E}_{32}(t)\,dt\right)
\right],
\nonumber\\
&\simeq&\exp\left[ -\,\mbox{\rm Im}\left(
\int_{\xi=0}^{\xi=1}\,
\sqrt{2}\,\Delta E_{31}\,
\chi\sqrt{1-\xi^2}
\,\frac{1}{|d\Lambda/dt|}d\Lambda\right)
\right],
\nonumber\\
&\simeq&\exp\left[ -
2\,\Delta E_{31}^2\,
\chi^2\,\mbox{\rm Im}
\left(\int_0^1\,
\,\frac{\sqrt{1-\xi^2}}{|d\Lambda/dt|} i d\xi\right)
\right],
\nonumber\\
&=&\exp\left[ -
2\,\Delta E_{31}^2\chi^2\,\int_0^1\,
\,\frac{\sqrt{1-\xi^2}}{|d\Lambda/dt|} d\xi
\right],
\nonumber
\end{eqnarray}
where we have used the fact
$d\Lambda=d(\Delta E_{31}\sqrt{u/3})=(\Delta
E_{31}/\sqrt{3})du/(2\sqrt{u})
\simeq (\Delta E_{31}/6)du$
$=6\sqrt{2}\Delta E_{31}\chi id\xi$.
For simplicity we assume that
$|d\Lambda/dt|=$ constant,
and we obtain Eq.\,(\ref{phb}).

\section{
The calculations of the effective mixing angle and
its derivative
in the case with large magnetic moments and a magnetic
field
\label{appendixh}}
To evaluate the derivative of $\tilde{\psi}_{23}$
we need to calculate not only the derivative of
$\{({\cal M})^j\}_{\tau\tau}$ which appear
in the KTY formula (\ref{solx}) but also the derivative of
the eigenvalues $\tilde{E}_j$.
The eigenvalues $\tilde{E}_j~(j=1,2,3)$ in Eq.\,(\ref{tj})
are written as
\begin{eqnarray}
\tilde{E}_j=2R\cos(\varphi+\frac{2j}{3}\pi)
=\frac{\Delta E_{31}}{3}\sqrt{1+u}\,
\cos(\varphi+\frac{2j}{3}\pi),
\nonumber
\end{eqnarray}
and, as we will see later, the derivative of $\varphi$ with respect
to $u$ vanishes at the level-crossing $u=u_0$.  To simplify the
calculations, therefore,
we introduce the normalized eigenvalues:
\begin{eqnarray}
\tilde{e}_j&\equiv&
\frac{\tilde{E}_j}{R}=2\cos(\varphi+\frac{2j}{3}\pi).
\label{tildee}
\end{eqnarray}
The normalized eigenvalues $\tilde{e}_j$ are
convenient when we compute the derivative
of $\tilde{\psi}_{23}$, because the derivative
of $\tilde{e}_j$ with respect to $u$ vanishes
at the level-crossing $u=u_0$.  So we express
the quantities, which are necessary to obtain
$\tilde{\psi}_{23}$, in terms of $\tilde{e}_j$:
\begin{eqnarray}
\tilde{X}_2^{\tau\tau}&=&\frac{1}{\Delta \tilde{e}_{32}\Delta \tilde{e}_{12}}
\left\{\tilde{e}_3\tilde{e}_1-(\tilde{e}_3+\tilde{e}_1)y_2^{\tau\tau}
+y_3^{\tau\tau}
\right\},
\nonumber\\
\tilde{X}_3^{\tau\tau}&=&\frac{1}{\Delta \tilde{e}_{31}\Delta \tilde{e}_{32}}
\left\{\tilde{e}_1\tilde{e}_2-(\tilde{e}_1+\tilde{e}_2)y_2^{\tau\tau}
+y_3^{\tau\tau}
\right\},
\nonumber\\
\tan^2\tilde{\psi}_{23}=
\frac{\tilde{X}_2^{\tau\tau}}
{\tilde{X}_3^{\tau\tau}}&=&
\frac{\Delta \tilde{e}_{31}}{\Delta \tilde{e}_{12}}\cdot
\frac{\tilde{e}_3\tilde{e}_1-(\tilde{e}_3+\tilde{e}_1)y_2^{\tau\tau}
+y_3^{\tau\tau}}{\tilde{e}_1\tilde{e}_2-(\tilde{e}_1+\tilde{e}_2)y_2^{\tau\tau}
+y_3^{\tau\tau}},
\label{psi23}
\end{eqnarray}
where we have defined
\begin{eqnarray}
\Delta\tilde{e}_{jk}&\equiv& \tilde{e}_j-\tilde{e}_k,
\nonumber\\
y_j^{\tau\tau}&\equiv&\frac{\left({\cal M}^{j-1}\right)_{\tau\tau}}
{R^{j-1}}.~(j=2,3)
\nonumber
\end{eqnarray}
The matrix elements of ${\cal M}$ and ${\cal M}^2$
can be calculated from Eq.\,(\ref{m0}) as follows:
\begin{eqnarray}
{\cal M}&=&
\left(
\begin{array}{ccc}
-\frac{1}{3}\Delta E_{31}&
-i\Lambda\cos\chi&
0\cr
i\Lambda\cos\chi&
-\frac{1}{3}\Delta E_{31}&
-i\Lambda\sin\chi\cr
0&
i\Lambda\sin\chi&
\frac{2}{3}\Delta E_{31}
\end{array}\right),
\nonumber\\
{\cal M}^2&=&
\left(
\begin{array}{ccc}
\frac{1}{9}\Delta E_{31}^2+\Lambda^2\cos^2\chi&
\frac{2}{3}i\Lambda\Delta E_{31}\cos\chi&
-\frac{1}{2}\Lambda^2\sin2\chi\cr
-\frac{2}{3}i\Lambda\Delta E_{31}\cos\chi&
\frac{1}{9}\Delta E_{31}^2+\Lambda^2&
-\frac{i}{3}\Lambda\Delta E_{31}\sin\chi\cr
-\frac{1}{2}\Lambda^2\sin2\chi&
\frac{i}{3}\Lambda\Delta E_{31}\sin\chi&
\frac{4}{9}\Delta E_{31}^2+\Lambda^2\sin^2\chi
\end{array}\right).
\nonumber
\end{eqnarray}

First of all, let us evaluate the value of $\tilde{\psi}_{23}$
itself at the level-crossing $u=u_0$.
At the level-crossing, from Eqs.\,(\ref{u0small}),
(\ref{phismall}) and (\ref{tildee}), we get
\begin{eqnarray}
\left.\tilde{e}_1\right|_{u=u_0}&\simeq&-2
\nonumber\\
\left.\tilde{e}_2\right|_{u=u_0}&\simeq&1+\frac{3}{2\sqrt{2}}\chi
\nonumber\\
\left.\tilde{e}_3\right|_{u=u_0}&\simeq&1-\frac{3}{2\sqrt{2}}\chi
\nonumber\\
\left.y_2^{\tau\tau}\right|_{u=u_0}&=&
\left.\frac{{\cal M}_{\tau\tau}}{R}\right|_{u=u_0}
=\left.\frac{(2/3)\Delta E_{31}}
{\sqrt{\Delta E_{31}^2/9+\Lambda^2/3}}\right|_{u=u_0}
=\left.\frac{2}{\sqrt{1+u}}\right|_{u=u_0}
\simeq 1
\nonumber\\
\left.y_3^{\tau\tau}\,\right|_{u=u_0}&=&
\left.\frac{({\cal M}^2)_{\tau\tau}}{R^2}\,\right|_{u=u_0}
=\left.\frac{(4/9)\Delta E_{31}^2+\Lambda^2\sin^2\chi}
{\Delta E_{31}^2/9+\Lambda^2/3}\,\right|_{u=u_0}
=\left.\frac{4+3u\sin^2\chi}{1+u}\,\right|_{u=u_0}
\simeq 1,
\nonumber
\end{eqnarray}
where we have used the fact
\begin{eqnarray}
R=
\sqrt{
\frac{\Delta E_{31}^2}{9}
+\frac{\Lambda^2}{3}}
=\frac{\Delta E_{31}}{3}
\sqrt{1+u},
\nonumber
\end{eqnarray}
and we have ignored terms of order $O(\chi^2)$.
Thus we observe that the effective mixing angle
$\tilde{\psi}_{23}$ is maximal at the level-crossing:
\begin{eqnarray}
\left.\tan^2\tilde{\psi}_{23}\right|_{u=u_0}
\simeq\frac{3}{-3}\cdot
\frac{-2(1-\frac{3}{2\sqrt{2}}\chi)+1+\frac{3}{2\sqrt{2}}\chi+1}
{-2(1+\frac{3}{2\sqrt{2}}\chi)+1-\frac{3}{2\sqrt{2}}\chi+1}
\simeq 1.
\nonumber
\end{eqnarray}
Hence $\tilde{\psi}_{23}$ is the appropriate
effective mixing angle to describe the jumping
factor $P_H$ as in the standard two flavor case.

Next, let us evaluate the derivative of $\tilde{\psi}_{23}$.
Since the derivative of $\varphi$
at the level-crossing vanishes
\begin{eqnarray}
\left.\frac{d\varphi}{dt}\right|_{u=u_0}
&=&\frac{du}{dt}\left.\frac{d\varphi}{du}\right|_{u=u_0}
\nonumber\\
&=&\frac{du}{dt}\left.\frac{d{\ }}{du}\left[\frac{1}{3}
\cos^{-1}\left\{\frac{1-Du}{(1+u)^{3/2}}\right\}\right]
\right|_{u=u_0}
\nonumber\\
&=&\frac{du}{dt}\left.\frac{-1}{3}
\left\{1-\frac{(1-Du)^2}{(1+u)^3}
\right\}^{-1/2}
\frac{D(u-u_0)}{2(1+u)^{5/2}}\right|_{u=u_0}=0,
\nonumber
\end{eqnarray}
the derivative of $\tilde{e}_j=2\cos(\varphi+2j\pi/3)$
at the level-crossing vanishes.
So the only terms which do not vanish upon
evaluating a derivative at the level-crossing $u=u_0$
in Eq.\,(\ref{psi23}) are
$y_2^{\tau\tau}$ and $y_3^{\tau\tau}$.
By taking the derivative of the logarithm of the
both hand sides of Eq.\,(\ref{psi23}), we get
at the level-crossing
\begin{eqnarray}
\left.\left(2\frac{d\tilde{\psi}_{23}}{du}\cdot
\frac{2}{\sin2\tilde{\psi}_{23}}\right)\right|_{u=u_0}
&=&\left.\left\{
\frac{-(\tilde{e}_3+\tilde{e}_1)\frac{dy_2^{\tau\tau}}{du}
+\frac{dy_3^{\tau\tau}}{du}}
{\tilde{e}_3\tilde{e}_1-(\tilde{e}_3+\tilde{e}_1)y_2^{\tau\tau}
+y_3^{\tau\tau}}
-\frac{-(\tilde{e}_1+\tilde{e}_2)\frac{dy_2^{\tau\tau}}{du}
+\frac{dy_3^{\tau\tau}}{du}}
{\tilde{e}_1\tilde{e}_2-(\tilde{e}_1+\tilde{e}_2)y_2^{\tau\tau}
+y_3^{\tau\tau}}\right\}\right|_{u=u_0}
\nonumber\\
&=&\left[\frac{\Delta\tilde{e}_{32}}
{\{\tilde{e}_3\tilde{e}_1-(\tilde{e}_3+\tilde{e}_1)y_2^{\tau\tau}
+y_3^{\tau\tau}\}
\{\tilde{e}_1\tilde{e}_2-(\tilde{e}_1+\tilde{e}_2)y_2^{\tau\tau}
+y_3^{\tau\tau}\}}\right.
\nonumber\\
&{\ }&\times\left.\left.
\left\{\frac{dy_2^{\tau\tau}}{du}(\tilde{e}_1^2-y_3^{\tau\tau})
+\frac{dy_3^{\tau\tau}}{du}(y_2^{\tau\tau}-\tilde{e}_1)\right\}
\right]\right|_{u=u_0}.
\label{psi231}
\end{eqnarray}
Here we note
\begin{eqnarray}
\left.\frac{dy_2^{\tau\tau}}{du}\right|_{u=u_0}&=&
\left.\frac{d{\ }}{du}\left(\frac{3}{\Delta
E_{31}\sqrt{1+u}}\frac{2}{3}\Delta E_{31}\right)\right|_{u=u_0}
\nonumber\\
&=&\left.-\frac{1}{(1+u)^{3/2}}\right|_{u=u_0}
\simeq -\frac{1}{8}
\nonumber\\
\frac{dy_3^{\tau\tau}}{du}&=&
\left.\frac{d{\ }}{du}\left\{\frac{9}{\Delta
E_{31}^2(1+u)}\left(
\frac{4}{9}\Delta E_{31}^2+\frac{u}{3}\Delta E_{31}^2
\sin^2\chi\right)\right\}\right|_{u=u_0}
\nonumber\\
&\simeq&\left.-\frac{4}{(1+u)^2}\right|_{u=u_0}
\simeq -\frac{1}{4}
\nonumber
\end{eqnarray}
where we have ignored terms of order $O(\chi^2)$.
Taking into account $\sin2\tilde{\psi}_{23}\simeq1$,
Eq.\,(\ref{psi231}) gives
\begin{eqnarray}
\left.\frac{d\tilde{\psi}_{23}}{du}\right|_{u=u_0}
&=&\frac{1}{4}\frac{3\chi/\sqrt{2}}{(-9\chi/2\sqrt{2})(9\chi/2\sqrt{2})}
\left\{-\frac{1}{8}(2^2-1)-\frac{1}{4}(1+2)
\right\}=\frac{1}{12\sqrt{2}\chi}.
\nonumber
\end{eqnarray}
Assuming $|d\Lambda/dt|_{u=u_0}=$ constant,
we have
\begin{eqnarray}
\left.\frac{d\tilde{\psi}_{23}}{dt}\right|_{u=u_0}
=\left.\frac{du}{dt}\right|_{u=u_0}\cdot
\left.\frac{d\tilde{\psi}_{23}}{du}\right|_{u=u_0}
=\frac{|d\Lambda/dt|_{u=u_0}}{\Delta E_{31}}\frac{1}{2\sqrt{2}\chi}
\nonumber
\end{eqnarray}
On the other hand,
\begin{eqnarray}
\left.\Delta \tilde{E}_{32}\right|_{u=u_0}
=\left. 2\sqrt{3}R\sin\left(\frac{\pi}{3}-\varphi\right)\right|_{u=u_0}
\simeq
\sqrt{2}\Delta E_{31}\chi
\nonumber
\end{eqnarray}
We conclude, therefore,
that the exponent of the
jumping factor coincides with
$-\pi/2$ times the $\gamma$ factor
in the case of a linear potential ($F=1$):
\begin{eqnarray}
\gamma=\left.\frac{\Delta \tilde{E}_{32}}
{2|d\tilde{\psi}_{23}/dt|}\right|_{u=u_0}
\simeq\frac{2\Delta E_{31}^2\chi^2}
{|d\Lambda/dt|_{u=u_0}}
=-\frac{\log P_H}{\pi/2}.
\nonumber
\end{eqnarray}

\section*{Acknowledgments}
The author would like to thank Akira Shudo for discussions
on the Landau-Zener theory in various stages of this work.
This research was partly supported by a Grant-in-Aid for Scientific
Research of the Ministry of Education, Science and Culture, under Grant
No. 24540281 and No. 25105009.

\end{document}